\documentclass[aps,pre,twocolumn,showpacs,superscriptaddress,groupedaddress]{revtex4}  
\usepackage{graphicx}  
\usepackage{dcolumn}   
\usepackage{bm}        
\usepackage{amssymb}   
\usepackage{hyperref}
\usepackage{color}
\usepackage{amsmath}

\begin{document}

\widetext

\title{Evolutionary games on cycles with strong selection}

\author{P.~M.~Altrock}
\email{paltrock@jimmy.harvard.edu}
\affiliation{Program for Evolutionary Dynamics, Harvard University, Cambridge, MA}
\affiliation{Dana-Farber Cancer Institute, Boston, MA}
\affiliation{Harvard T.~H.~Chan School of Public Health, Boston, MA}

\author{A.~Traulsen}
\affiliation{Department of Evolutionary Theory, Max Planck Institute for Evolutionary Biology, Pl\"on, Germany}

\author{M.~A.~Nowak}
\affiliation{Program for Evolutionary Dynamics, Harvard University, Cambridge, MA}

\date{\today}

\begin{abstract}
Evolutionary games on graphs describe how strategic interactions and population structure determine evolutionary success, quantified by the probability that a single mutant takes over a population. 
Graph structures, compared to the well-mixed case, can act as amplifiers or suppressors of selection by increasing or decreasing the fixation probability of a beneficial mutant. 
Properties of the associated mean fixation times can be more intricate, especially when selection is strong. 
The intuition is that fixation of a beneficial mutant happens fast (in a dominance game), that fixation takes very long (in a coexistence game), and that strong selection eliminates demographic noise. Here we show that these intuitions can be misleading in structured populations.  
We analyze mean fixation times on the cycle graph under strong frequency-dependent selection for two different microscopic evolutionary update rules (death-birth and birth-death). 
We establish exact analytical results for fixation times under strong selection, and show that there are coexistence games in which fixation occurs in time polynomial in population size. 
Depending on the underlying game, we observe inherence of demographic noise even under strong selection, if the process is driven by random death before selection for birth of an offspring (death-birth update). 
In contrast, if selection for an offspring occurs before random removal (birth-death update), strong selection can remove demographic noise almost entirely. 
\end{abstract}
\pacs{02.50.Ga, 87.23.Kg, 87.23.Cc, 87.18.-h}
\maketitle

\section{\label{sec:intro}Introduction}

Evolutionary game theory models Darwinian selection among genetically hard-wired strategic traits or behaviors in a population \cite{maynard-smith:Nature:1973,hofbauer:book:1988}. Often the interaction between behaviors is cast into an evolutionary game, and the performance in this evolutionary game determines the rate at which strategies spread. As payoffs from the game are mapped to fitness, i.e.~the expected number of offspring in the near future, more successful behaviors have a higher tendency to spread in the population. In infinitely large populations this spreading of successful behaviors due to Darwinian selection is described by the deterministic replicator dynamics \cite{taylor:MB:1978,hofbauer:JTB:1979,zeeman:LN:1980,ohtsuki:JTB:2006b,ohtsuki:JTB:2008}. In finite populations fluctuations cannot be neglected and the evolutionary dynamics become stochastic \cite{fogel:EM:1998,nowak:Nature:2004,traulsen:PRL:2005,hilbe:BMB:2011,black:TREE:2012}. A parameter that governs the interplay between the determinism of selection and intrinsic stochasticity in finite populations is the strength of selection \cite{altrock:PRE:2010}. Neutral evolution emerges in absence of selective differences. If selection acts, the stochastic evolutionary dynamics become payoff dependent, which can be the same in each state (constant selection) or entirely state dependent (frequency-dependent selection), whereby the state is defined as the number of mutants. In the case of frequency-dependent selection the probability that one strategy replaces another can be fairly complicated. In particular, structure of the population itself influences the evolutionary game and the potential success of strategic behaviors \cite{lieberman:Nature:2005,broom:PRSA:2008,broom:PRSA:2010,hindersin:JRSI:2014,kaveh:JRSOS:2015}.

In stochastic evolutionary game dynamics the event of interest is fixation of a mutant \cite{nowak:book:2006}. Two quantities have been of special interest: the fixation probability and the expected fixation time \cite{maruyama:GR:1970,nei:Genetics:1973,slatkin:Evolution:1981,taylor:JTB:2006,altrock:NJP:2009,altrock:PRE:2010}. Fixation probabilities have served as the gauge whether a graph can be an amplifier of suppressor of selection \cite{lieberman:Nature:2005,hindersin:PlosCB:2015}. An open problem is the general quantification of fixation times in graph structured populations. Fixation times quantify the expected time new traits need to take over the population. For constant selection, recent findings have established that evolution can slow down substantially in populations where selection is amplified \cite{frean:PRSB:2013}, and that there are no obvious relations between fixation probabilities and fixation times on graphs \cite{hindersin:JRSI:2014}. In addition,  structured population dynamics may be different if selection occurs before or after random death of individuals \cite{ohtsuki:Nature:2006,hindersin:PlosCB:2015}. Here we seek to start closing this gap using analytical procedures in an evolutionary game between resident strategy and a mutant strategy. To this end, we consider dynamics on the simplest structure, in which exactly one individual occupies a node on an undirected cycle graph \cite{ohtsuki:PRSB:2006}, and focus on strong selection \cite{altrock:PRE:2010}.

This manuscript is organized in the following way. First, we introduce the stochastic evolutionary dynamics. We review the well-mixed population and discuss the analytical expressions for fixation probabilities, sojourn times and fixation times. Then, we introduce the transition probabilities of birth-death and death-birth updating on cycle graphs. In the results section, we consider neutral evolution and briefly address constant selection before we turn to strong frequency dependent selection. Then, we discuss standard cases of two player-two strategy games between a mutant strategy $A$ and a resident strategy $B$, given by the payoff matrix
\begin{eqnarray}
 \begin{array}{c|cc}
  & A & B \\ \hline
  A & a & b \\
  B & c & d
 \end{array}\label{eq:payoffmatrix}
\end{eqnarray}
We consider strategic dominance games of the mutant strategy in which $A$ always has a higher payoff ($a>c$, $b>d$), coordination games in which both mutants and residents receive highest payoffs when interacting with their own types ($a>c$, $b<d$), and coexistence games in which both mutants and residents receive highest payoffs when interacting with the other respective strategy ($a<c$, $b>d$).  
As particular examples, we discuss the Prisoner's Dilemma (where defection dominates cooperation) an the and the Snowdrift Game (where defection and cooperation can coexists).

\section{\label{sec:EGD}Evolutionary game dynamics}

First we describe the discrete-time Markov chain model resulting from subsequent birth and death events in a population of finite fixed size. Two key assumptions are that we start with a single mutant individual and that no further mutations occur. 
Thus, on a cycle graph, the mutant population can only grow as a cluster. 
We also make the  assumption that replacement graph and interaction graph are identical \citep{ohtsuki:PRSB:2006}. 
The resulting Markov chain eventually gets absorbed either of the boundary state of no or all mutants. The population size is $N$ and we denote the evolutionary transition probabilities by $T^{i\pm}$. Here, $i$ is the number of mutant individuals that at any time can increase or decrease by exactly one. The process stays in state $i$ with probability $1-T^{i+}-T^{i-}$. 

We can then examine the fixation probability of a group of $i$ type $A$ individuals, $\phi^i$, as well as other quantities of interest without specifying the transition probabilities. 
The fixation probability follows from solving the backward Kolmogorov equation $\phi^i=T^{i+}\phi^{i+1}+T^{i-}\phi^{i-1}+(1-T^{i+}-T^{i-})\phi^i$ recursively \citep[see e.g.][]{karlin:book:1975,nowak:book:2006,altrock:PRE:2010}
and is given by
\begin{eqnarray}\label{eq:dBRing_FP01}
\begin{split}
\phi^i = \frac{1+\sum_{k=1}^{i-1}\prod_{l=1}^k\frac{T^{l-}}{T^{l+}}}{1+\sum_{k=1}^{N-1}\prod_{l=1}^k\frac{T^{l-}}{T^{l+}}}
\end{split}
\end{eqnarray}
where only the ratios of transition probabilities enter. 

To characterize the expected time scale of the evolutionary process, one can consider two different quantities. First, the mean unconditional fixation time describes the expected time the process takes to reach either extinction or fixation of the mutant, which occurs with probability one. Second, the mean conditional fixation time describes the expected number of time steps the process takes to fixation of the mutant, which occurs with probability $\phi^1$. One way to derive an expression for mean fixation times is to think about the expected time spent in each intermediate state (including waiting times) $j=1,2,\dots,N\!-\!2,N\!-\!1$, which are called sojourn times. 
The sojourn time of a particular state $j$ can be found by considering its escape process.
Say that at time $t^0$, the process is in state $j$. 
Then, the probability that it either stays or ever returns to that state $j$ (denoted as the super-script) is given by 
\begin{eqnarray}\label{eq:sojourn_01}
r^j = (1-T^{j+}-T^{j-}) + T^{j+}\,\phi^{j+1,j} + T^{j-}\,\phi^{j-1,j}
\end{eqnarray}
Here, $\phi^{k,l}$ is the probability to start in state $k$ and ever return to state $l$ ($\phi^{k,k}=1$) \citep{altrock:JTB:2012}, which is not conditioned on fixation. The conditional probability to start in state $i$, return a positive number of time steps $t$ to that state, but then escape is given by 
\begin{eqnarray}\label{eq:sojourn_01x}
\phi^{i,j}\,(r^j)^{t-1}\,(1-r^j)
\end{eqnarray}
The first factor of Eq.~(\ref{eq:sojourn_01x}) describes the probability to ever get from $i$ to $j$, the second factor describes recurrence such that the total time spent amounts to exactly $t$ time steps, and the third factor describes definite escape from state $j$. 

The mean sojourn time in state $j$, starting from one mutant, $i=1$, is thus given by the first moment of this conditional probability
\begin{eqnarray}\label{eq:sojourn_02}
t^{1,j} = \sum\limits_{t=1}^{\infty}\phi^{1,j}\,(r^j)^{t-1}\,(1-r^j) t
\end{eqnarray}
This geometric sum can be solved exactly, and inserting the definition of $r^j$ we obtain an exact expression for the mean sojourn time of state $j$, $t^{1,j} =\phi^{1,j}/(1-r^j)$. 
The mean unconditional fixation time is then given by the sum over all sojourn times \citep{ewens:book:2004,ohtsuki:PRSB:2006,altrock:JTB:2012}
\begin{eqnarray}\label{eq:sojourn_03}
\begin{split}
t^1 
=\sum\limits_{j=1}^{N-1}\frac{\phi^{1,j}}{T^{j+}(1-\phi^{j+1,j}) + T^{j-}(1-\phi^{j-1,j})}
\end{split}
\end{eqnarray}
The mean conditional fixation time can be found in a similar way, only resorting to conditional transition probabilities of the Markov process $\widetilde T^{i\to j}= \left(\phi^j/\phi^i\right)T^{i\to j}$ \citep{ewens:book:2004}, from which we find the probabilities to start in $i$ and ever visit $j$ under the condition of fixation in $N$, $\widetilde\phi^{i,j} = \left(\phi^j/\phi^i\right)\times\phi^{i,j}$. Consequently, the mean sojourn time in state $j$ conditioned on fixation, starting from a single mutant, can be expressed by the mean unconditional sojourn time 
\begin{eqnarray}\label{eq:sojourn_04}
\widetilde t^{1j} = \frac{\phi^j}{\phi^1}\,t^{1,j}
\end{eqnarray}
such that the mean conditional fixation time can be calculated by 
\begin{eqnarray}\label{eq:sojourn_04}
t^{1\to N} = \sum\limits_{j=1}^{N-1}\frac{\phi^j}{\phi^1}\,\frac{\phi^{1,j}}{T^{j+}(1-\phi^{j+1,j}) + T^{j-}(1-\phi^{j-1,j})}
\end{eqnarray}
whereby the stationary probabilities to ever go from state $i$ to state $j$, $\phi^{i,j}$ are derived in the book by Ewens \cite{ewens:book:2004}. We repeat them in our Appendix, Eqs. (\ref{eq:sojourn_05}) and (\ref{eq:sojourn_06}).

\subsection{Well-mixed population}

The reference case for evolutionary game dynamics is the well-mixed population \cite{nowak:Science:2004}. In the well-mixed population, an expected payoff is calculated taking into account interactions between all individuals. This is equivalent to a fully connected unweighted graph. Here, we briefly recall the properties of the Moran process of frequency dependent selection in well-mixed populations. Formally, the Moran process is introduced as a birth-death process, but in a well-mixed population, the ordering of a fitness-proportional birth and a random death event does not have any influence on the dynamics as long as we include self replacement, which is commonly assumed \cite{hindersin:PlosCB:2015}. 

The well-mixed population assumes that there are interactions between all individuals, which lead to an average payoff, 
and in turn determines selection via a specific choice of payoff to fitness mapping \citep{wu:PRE:2010,wu:PlosCB:2013,wu:NJP:2015}.
In this paper, we focus on an exponential payoff to fitness mapping \citep{traulsen:bmb:2008}. 
If $i$ and $N-i$ are the numbers of $A$ and $B$ individuals, the expected payoff of any $A$ individual is given by $\pi_A=a\,(i-1)/(N-1)+b\,(N-i)/(N-1)$. The expected payoff of any $B$ individual is given by $\pi_B=c\,i/(N-1)+d\,(N-i-1)/(N-1)$. 
Then, the Moran process is a Markov chain with transition probabilities to neighboring states given by
\begin{eqnarray}
	T_{\text{wm}}^{i+} &=& \frac{i\,\text{e}^{\beta\pi_A}}{i\,\text{e}^{\beta\,\pi_A}+(N-i)\text{e}^{\beta\,\pi_B}}\frac{N-i}{N}	\label{eq:Moran_01a}, \\
	T_{\text{wm}}^{i-} &=&	 \frac{(N-i)\text{e}^{\beta\,\pi_B}}{i\,\text{e}^{\beta\,\pi_A}+(N-i)\text{e}^{\beta\,\pi_B}}\frac{i}{N}	\label{eq:Moran_01b}.
\end{eqnarray}
These transition probabilities are non-linear functions of the number of mutant individuals. 
For most types of game in well-mixed populations, the states between all-mutant and all-resident have different probabilities to increase or decrease the number of mutants. 

\begin{figure}[h]
\centering
\includegraphics[width=1.00\columnwidth]{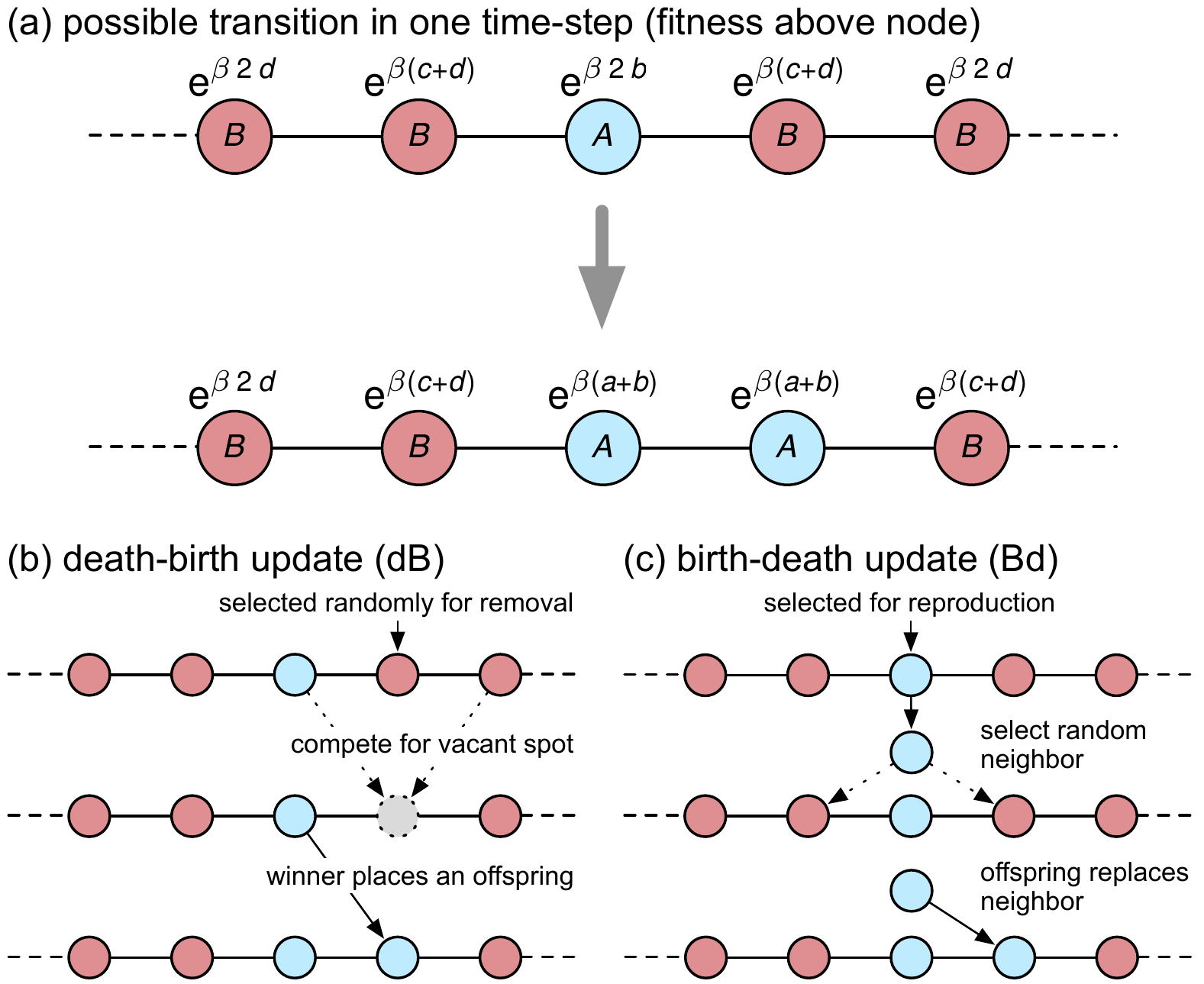}
\caption{\label{fig:Fig01}
(color online) \textbf{Example of a possible transition.}
During a single time step, one possible event is that the number of mutants ($A$) increases by one at the expense of one resident ($B$) (shown in a) (otherwise it could decrease by one, or stay constant). These transitions are driven by the fitness values of the nodes involved. We use the exponential payoff to fitness conversion with selection intensity $\beta$, shown for each node above. The fitness from a mutant-mutant interaction is $\text{exp}(\beta\,a)$, from a the mutant-resident interaction it is $\text{exp}(\beta\,b)$, and from a resident-resident interaction it is $\text{exp}(\beta\,d)$. Transitions from $i$ to $i+1$ mutants can be described by the death-birth update rule (b), Eqs.~(\ref{eq:dBRing_p})-(\ref{eq:dBRing_m}), or by the birth-death update rule (c), Eqs.~(\ref{eq:BdRing_p})-(\ref{eq:BdRing_m}). These basic local rules matter for the time scale of the process of fixation (takeover or extinction of mutants). 
}
\end{figure}

There are two popular mechanisms often used to describe evolutionary dynamics on graph structured populations \citep{lieberman:Nature:2005,antal:PRL:2006,ohtsuki:Nature:2006,szabo:PR:2007,zukewich:PlosOne:2013}. First, in the death-birth process, there is random death of an individual and subsequent selection among its neighbors for filling the vacant spot. Thus, competition is only among individuals of the immediate neighborhood of the vacant spot. Second, in the Birth-death process, there is selection for birth of an identical offspring within the entire population, before random death of a neighbor of the reproducing individual occurs. Hence, the death-birth process models random death which precedes local competition, whereas the birth-death process models global completion which precedes random death. 
The basic difference between the two update mechanism on the cycle are depicted in Figure \ref{fig:Fig01}, and described in more detail in the following.

\subsection{Death-birth process on a cycle}

The death-birth update on a cycle works as follows. Each node of the graph represents one individual. First, all individuals play the evolutionary game with their two neighbors. Then a random individual is selected for removal. The neighbors of this empty spot then compete for placing an identical offspring. The number of mutants only changes when random death occurs at the boundary, as mutants grow in a cluster. There are two sites that can be chosen for random death at the boundary of residents and mutants, which occurs with probability $2/N$. Competition among the two neighbors of the vacant spot then leads to a probability that a mutant fills the spot. Let $f_A$ and $f_B$ be the fitness values of the two neighbors of the vacant spot. Then, the probability that the $A$ individual's offspring occupies the vacant spot is $f_A/(f_A+f_B)=1/(1+f_B/f_A)$, resulting in the transition probabilities
\begin{eqnarray}\label{eq:dBRing_p}
\begin{split}
T_{\text{dB}}^{i+} = 
\begin{cases}
\frac{2}{N}\frac{1}{1+\text{e}^{-\beta(2b-2d)}}\,\,\,&\text{if}\,\,\,i=1\\
\frac{2}{N}\frac{1}{1+\text{e}^{-\beta(a+b-2d)}}\,\,\,&\text{if}\,\,\,1<i<N-2\\
\frac{2}{N}\frac{1}{1+\text{e}^{-\beta(a+b-c-d)}}\,\,\,&\text{if}\,\,\,i=N-2\\
\frac{1}{N}\,\,\,&\text{if}\,\,\,i=N-1\\
0\,\,\,&\text{if}\,\,\,i=0,N
\end{cases}
\end{split}
\end{eqnarray}
and 
\begin{eqnarray}\label{eq:dBRing_m}
\begin{split}
T_{\text{dB}}^{i-} = 
\begin{cases}
\frac{1}{N}\,\,\,&\text{if}\,\,\,i=1\\
\frac{2}{N}\frac{1}{1+\text{e}^{+\beta(a+b-c-d)}}\,\,\,&\text{if}\,\,\,i=2\\
\frac{2}{N}\frac{1}{1+\text{e}^{+\beta(2a-c-d)}}\,\,\,&\text{if}\,\,\,2<i<N-1\\
\frac{2}{N}\frac{1}{1+\text{e}^{+\beta(2a-2c)}}\,\,\,&\text{if}\,\,\,i=N-1\\
0\,\,\,&\text{if}\,\,\,i=0,N
\end{cases}
\end{split}
\end{eqnarray}
The states with one mutant individual or one resident individual are special in the sense that there is no competition, only random selection for removing that last individual. Interestingly, for $2<i<N-2$, the expansion of a cluster of $A$ players does not depend on the payoff parameter $c$. This makes sense as the only individual affected by $c$ had to be subject to random death before selection. Similarly, the expansion of a $B$-cluster does not involve the payoff parameter $b$. Note also that since all transition probabilities are determined by payoff differences, this process is invariant under adding a constant to each payoff value, and multiplication of the payoff matrix with a positive real number changes the strength of selection.

\subsection{Birth-death process on a cycle}

For the frequency dependent birth-death ($\text{Bd}$) update on a cycle, with exponential payoff to fitness mapping, we find more complicated transition probabilities. The transition from one mutant to a cluster of two mutants occurs with probability
\begin{eqnarray}\label{eq:dBRing_p01}
T_{\text{Bd}}^{1+}= 
\frac{\text{e}^{\beta\,2 b}}{\text{e}^{ \beta\,2 b}+2 \text{e}^{\beta(c+d)}+(N-3) \text{e}^{\beta\,2d}}	
\end{eqnarray}
and the extinction of a single mutant occurs with probability
\begin{eqnarray}\label{eq:dBRing_p02}
T_{\text{Bd}}^{1-}= 
\frac{\text{e}^{\beta  (c+d)}}{\text{e}^{\beta\,2 b}+2 \text{e}^{\beta  (c+d)}+(N-3) \text{e}^{\beta\,2 d}}
\end{eqnarray}
In similar fashion we find
\begin{eqnarray}\label{eq:BdRing_p}
\begin{split}
T_{\text{Bd}}^{i+}= 
\begin{cases}
\frac{\text{e}^{\beta  (a+b)}}{
F_i
}		\,\,\,&\text{if}\,\,\,1<i<N-1\\
\frac{\text{e}^{\beta  (a+b)}}{
F_{N-1}
}		\,\,\,&\text{if}\,\,\,i=N-1\\
0\,\,\,&\text{if}\,\,\,i=0,N
\end{cases}
\end{split}
\end{eqnarray}
and 
\begin{eqnarray}\label{eq:BdRing_m}
\begin{split}
T_{\text{Bd}}^{i-} = 
\begin{cases}
\frac{\text{e}^{\beta  (c+d)}}{
F_i
}		\,\,\,&\text{if}\,\,\,1<i<N-1\\
\frac{\text{e}^{\beta\,2 c}}{
F_{N-1}
}		\,\,\,&\text{if}\,\,\,i=N-1\\
0		\,\,\,&\text{if}\,\,\,i=0,N
\end{cases}
\end{split}
\end{eqnarray}
where we used the abbreviations $F_i$ for the total fitness of the population in state $i$, see Appendix \ref{sec:Bdtransitions}. Again, the birth-death process is invariant under adding a constant to the payoff values. 
In this case, there are only three states for which the ratio of transition probabilities $t_{\text{Bd}}^{i-} / t_{\text{Bd}}^{i+}$ deviates from the constant value $e^{-\beta(a+b-c-d)}$, we obtain a different value only for $i=1,2,N-2,$ and $N-1$.

\section{\label{sec:results}Results}

\subsection{Neutral evolution}

Neutral evolution means that the probability to increase or to decrease the number of mutants on the cycle are always the same, irrespective of the number of mutant individuals. On a cycle, this probability is the same for any number of mutants as long as neither mutant or resident is extinct. The probably to increase or decrease the number of mutants also becomes independent of the specific update mechanism used and simply amounts to $1/N$. Under any of the two update rules we thus find  
\begin{eqnarray}\label{eq:neutral_01}
\begin{split}
\phi^i_{\text{dB}} =\phi^i_{\text{Bd}} = \frac{i}{N}
\end{split}
\end{eqnarray}
for the fixation probability starting from $i$ mutants. For the fixation times stating from one mutant we obtain
\begin{eqnarray}
t^1_{\text{dB}} = t^1_{\text{Bd}} &=& \frac{1}{2}(N-1)N \label{eq:neutral_02}
\end{eqnarray}
for the mean unconditional fixation times, and
\begin{eqnarray}
t^{1\to N}_{\text{dB}}=t^{1\to N}_{\text{Bd}} &=& \frac{1}{6}(N-1)N(N+1) \label{eq:neutral_03}
\end{eqnarray}
for the mean conditional fixation times. 
For the well-mixed population the the mean unconditional fixation time of the neutral process is $t^1_{\text{wm}}=N\,H_{N-1}$, where $H_k=1+1/2+\dots+1/k$ is the harmonic number, which increases logarithmically for large $k$. Thus, fixation or loss is slower on the cycle graph, where it scales as $\sim N^2$. Also the mean conditional fixation time on the cycle ($\sim N^3$) is much longer than 
in a well mixed population, where it only scales quadratically in $N$,  $t^{1\to N}_\text{wm}=N(N-1)$. Often, the times under neutral evolution set the reference against which the mean times under selection are measured \citep{altrock:PRE:2010,altrock:JTB:2012,hindersin:PlosCB:2015}. 

\subsection{Constant selection}

For constant selection we assume $f_A=\text{e}^{\beta\,r}$ and $f_B=1$. In this case, the death-birth process and the birth-death process are inherently different (see Appendix). The death-birth process under strong constant selection in favor of the mutant, $\beta\,r \to \infty$, leads to mean sojourn times that are constant proportional to $N$. The birth-death process under strong constant selection leads to mean sojourn times in state $j$ that are equal to $j$. Ultimately, this leads to
\begin{eqnarray}
t^{1\to N}_\text{dB}=t^1_\text{dB} &\to& \frac{N^2}{3} \label{eq:constant01}\\
t^{1\to N}_\text{Bd}=t^1_\text{Bd} &\to& N\frac{N-1}{2} \label{eq:constant02},
\end{eqnarray}
where the conditional and unconditional fixation times are identical because the fixation probability quickly converges to 1. We can also make a quick comparison of the Moran process on the ring (the Bd process) and the Moran process in a well-mixed population. In Appendix \ref{sec:constantExpansion} calculate approximations for these two strongly related processes for finite but large $r$, resulting in the mean conditional fixation times $t^{1\to N}_{\text{Bd}}\approx N(N-1)/2+N(N-2)\text{e}^{-\beta\,r}$ and $t^{1\to N}_{\text{WM}}\approx N\,H_{N-1}+2\,N\,H_{N-2}\,\text{e}^{-\beta\,r}$. These relations are able to describe how the mean fixation times approach the value of the strong selection limit value for any population larger than $N=2$.

\subsection{Strong frequency-dependent selection}

Now, we discuss the semi-analytical solutions given by Equations (\ref{eq:sojourn_03}) and (\ref{eq:sojourn_04}), for frequency-dependent selection. We focus on the limiting case of strong selection, $\beta\to\infty$, to obtain further analytical insights. These insights can guide our intuition as to how strong frequency dependent selection and spatial structure impact times to extinction or fixation in structured populations of finite size. 

We now quantify how the different update rules on the cycle behave in the strong selection limit in order to develop an intuition for the limiting behavior of fixation probabilities, as well as mean fixation times. We focus on non-trivial sets in payoff-space, e.g.~$a+b>c+d$, and exclude special cases which are of measure zero in parameter space, e.g.~$b=d$. We start with the dB update on the cycle. The limiting cases of $\beta \to \infty$ for the transition probabilities to increase the number of $A$ individuals are given by
\begin{eqnarray}
		T_{\text{dB}}^{1+} &\to& 
		\begin{cases}
			\frac{2}{N}\,\,\,\text{if}\,\,\,b>d\\
			0\,\,\,\text{if}\,\,\,b<d
		\end{cases} \label{eq:dBStrong_01a}\\
		T_{\text{dB}}^{i+}&\to& 
		\begin{cases}
			\frac{2}{N}\,\,\,\text{if}\,\,\,a+b>2d\\
			0\,\,\,\text{if}\,\,\,a+b<2d
		\end{cases} \label{eq:dBStrong_01b}\\
		T_{\text{dB}}^{(N-2)+} &\to& 
		\begin{cases}
			\frac{2}{N}\,\,\,\text{if}\,\,\,a+b>c+d\\
			0\,\,\,\text{if}\,\,\,a+b<c+d
		\end{cases} \label{eq:dBStrong_01c}\\
		T_{\text{dB}}^{(N-1)+} &\to& \frac{1}{N}\label{eq:dBStrong_01d}
\end{eqnarray}
In the same way, we obtain the limiting cases of the transition probabilities to decrease the number of $A$ individuals:
\begin{eqnarray}
		T_{\text{dB}}^{1-} &\to& \frac{1}{N} \label{eq:dBStrong_02a}\\
		T_{\text{dB}}^{2-} &\to& 
		\begin{cases}
			\frac{2}{N}\,\,\,\text{if}\,\,\,a+b<c+d\\
			0\,\,\,\text{if}\,\,\,a+b>c+d
		\end{cases} \label{eq:dBStrong_02b}\\
		T_{\text{dB}}^{i-} &\to& 
		\begin{cases}
			\frac{2}{N}\,\,\,\text{if}\,\,\,2a<c+d\\
			0\,\,\,\text{if}\,\,\,2a>c+d
		\end{cases} \label{eq:dBStrong_02c}\\
		T_{\text{dB}}^{(N-1)-} &\to& 
		\begin{cases}
			\frac{2}{N}\,\,\,\text{if}\,\,\,a<c\\
			0\,\,\,\text{if}\,\,\,a>c
		\end{cases} \label{eq:dBStrong_02d}
\end{eqnarray}
 where again, the last step before $B$ fixes in the population is independent of the game. 
 
The spread of a single $A$ mutant is impossible under strong selection if the payoff of $A$ against $B$ is lower than the payoff of $B$ against itself. However, the loss of the single $A$ mutant occurs with an expected waiting time proportional to the size of the population. Fixation of $A$ can only occur under strong selection if this initial step is possible ($b>d$) and if additionally $a+b>2d$, as well as $a+b>c+d$. 

If there is an unstable mixed Nash equilibrium in the game (which is then a coordination game), fixation of $A$ is only possible in a subset of all games. Generally, under the dB process fixation and extinction may still take long in large populations, as the non-vanishing transition probabilities are proportional to $N^{-1}$.

In contrast, the transitions rates of the Bd process (evolutionary Moran process on graphs \citep{ohtsuki:Nature:2006}) can become independent of the population size, 
The limiting cases for the transition probabilities are given by the following conditions, where it is important to note that for non-trivial transition probabilities, {\em all} the payoff relations have to be fulfilled, 
\begin{eqnarray}
		T_{\text{Bd}}^{1+} \!&\to&\! 
		\begin{cases}
			1\,\,\,\text{if}\,\,\,2b>c+d, b>d\\
			0\,\,\,\text{if}\,\,\,2b<c+d\,\text{or}\,b<d
		\end{cases} \label{eq:BdStrong_01a}\\
		T_{\text{Bd}}^{i+} \!&\to&\! 
		\begin{cases}
			\frac{1}{2}\,\,\,\text{if}\,\,\,b>a, a+b>c+d, a+b>2d\\
			0\,\,\text{if}\,\,b\!<\!a, a\!+\!b\!<\!c\!+\!d,\,\text{or}\,a\!+\!b\!<\!2d
		\end{cases} \label{eq:BdStrong_01b}\\
		T_{\text{Bd}}^{(N-1)+} \!&\to&\! 
		\begin{cases}
			\frac{1}{2}\,\,\,\text{if}\,\,\,b>a, a+b>2c\\
			0\,\,\,\text{if}\,\,\,b<a \,\text{or}\, a+b<2c 
		\end{cases} \label{eq:BdStrong_01c}
\end{eqnarray}
and
\begin{eqnarray}
		T_{\text{Bd}}^{1-} \!&\to&\! 
		\begin{cases}
			\frac{1}{2}\,\,\,\text{if}\,\,\,c>d\,\text{and}\,c+d>2b\\
			0\,\,\,\text{if}\,\,\,c<d\,\text{or}\,c+d<2b
		\end{cases} \label{eq:BdStrong_02a}\\
		T_{\text{Bd}}^{i-} \!&\to&\! 
		\begin{cases}
			\frac{1}{2}\,\,\,\text{if}\,\,\, c+d>a+b, c+d>2b, c>d\\
			0\,\,\,\text{if}\,\,c\!+\!d\!<\!a\!+\!b, c\!+\!d\!<\!2b,\,\text{or}\,c\!<\!d
		\end{cases} \label{eq:BdStrong_02b}\\
		T_{\text{Bd}}^{(N-1)-} \!&\to&\! 
		\begin{cases}
			1\,\,\,\text{if}\,\,\,2c>a+b, c>a\\
			0\,\,\,\text{if either}\,\,2c<a+b, c<a 
		\end{cases} \label{eq:BdStrong_02c}
\end{eqnarray}
Hence, the last transition before extinction or fixation of $A$ can become entirely deterministic. All other transitions have a non-trivial limiting case of $1/2$, independent of the size of the population, which is due to the fact that the selection (birth) step comes first, which deterministically selects the best performing individual(s). This selection step is then followed by the death step, which can only select one of the two neighbors of the parent. Thus, under strong selection, fixation in a Bd process can occur at a much faster rate than fixation in a dB process. In addition, it is possible that strong selection on the cycle becomes static; neither increase or decrease of the mutant strategy $A$ are possible. 

For comparison we again consult the well-mixed population, where, based on Eqs.~(\ref{eq:Moran_01a}) and (\ref{eq:Moran_01b}), the dynamics always depends on the payoff in the form of relations between expected payoffs $\pi_A$, $\pi_B$. As a consequence, the strong selection limiting cases in the well-mixed population are 
\begin{eqnarray}
T_{\text{wm}}^{i+} &\to& \begin{cases}
			\frac{N-i}{N} \,\,\,\text{if}\,\,\,a\frac{i}{N-i}+b>c\frac{i}{N-i}+d\\
			0\,\,\,\text{if}\,\,\,a\frac{i}{N-i}+b<c\frac{i}{N-i}+d
		\end{cases} \label{eq:BdStrong_03a}\\
T_{\text{wm}}^{i-} &\to& \begin{cases}
			\frac{i}{N} \,\,\,\text{if}\,\,\,a\frac{i}{N-i}+b<c\frac{i}{N-i}+d\\
			0\,\,\,\text{if}\,\,\,a\frac{i}{N-i}+b>c\frac{i}{N-i}+d
		\end{cases}\label{eq:BdStrong_03b}
\end{eqnarray}
The dynamic time scales of fixation or extinction processes are generally population size dependent, but depend on the payoff matrix only implicitly. For sufficiently large but finite population size $N$, the mean unconditional fixation time of a strategic dominance game or a suitable coordination game is proportional to $N\,\text{log}[N-1]$ plus a constant (in fact Euler's constant \cite{graham:book:1994}), but generally diverges exponentially for coexistence games \cite{traulsen:PRE:2006c,altrock:JTB:2012}.

\section{\label{sec:discussion}Discussion}

\subsection{Strategic dominance}

We speak of strategic dominance if in any one shot interaction, $A$ does always better than $B$, i.e. $a>c$, $b>d$. In the well mixed population, this immediately leads to a relation for the transition rates for any strength of selection, $T_{\text{wm}}^{i+}\geq T_{\text{wm}}^{i-}$. In structured populations, additional conditions on the payoffs may be required for this to be fulfilled. 

For the {\bf death-birth process} the strong selection limiting case leads to the following payoff relations. If the three inequalities $a+b>2d$, $a+b>c+d$, and $2a>c+d$ hold, we obtain limiting cases of the transition rates and sojourn times,
\begin{eqnarray}
 \begin{array}{c|ccccccc}
    & 1 & 2 & \dots & i & \dots & N-2 & N-1  \\ \hline
  T_{\text{dB}}^{i+} & \frac{2}{N} & \frac{2}{N} & \dots & \frac{2}{N} & \dots & \frac{2}{N} & \frac{1}{N}\\
  T_{\text{dB}}^{i-}  & \frac{1}{N} & 0 & \dots & 0 & \dots & 0 & 0 \\
  t^{1 i}  & \frac{N}{3} & \frac{N}{3} & \dots & \frac{N}{3} & \dots & \frac{N}{3} & \frac{2N}{3}
 \end{array}\label{eq:Dominance_dB_04}
\end{eqnarray}
A single mutant has an extinction probability of only  one-third, but a probability to reach fixation of two-thirds: once the mutation spreads to two or more individuals, it is bound to take over with certainty. The fixation times can again be calculated by summation over the sojourn times, using Eqs.~(\ref{eq:sojourn_03}) and (\ref{eq:sojourn_04}), which leads to
\begin{eqnarray}
	t^1_\text{dB} \to \frac{N^2}{3}
\label{eq:Dominance_dB_05}
\end{eqnarray}
and
\begin{eqnarray}
	t^{1\to N}_\text{dB} \to \frac{N^2}{2}-\frac{N}{6}
\label{eq:Dominance_dB_06}
\end{eqnarray}
In both cases, the leading order term is quadratic in the population size $N$. 

The second case of non-vanishing fixation probability of a single mutant is when $2a < c+d$. Then, fixation of the dominant mutant takes considerably longer on average but the fixation probability of a single mutant is still $2/3$:
\begin{eqnarray}
 \begin{array}{c|ccccccc}
    & 1 & 2 & \dots & i & \dots & N-2 & N-1  \\ \hline
  T_{\text{dB}}^{i+} & \frac{2}{N} & \frac{2}{N} & \dots & \frac{2}{N} & \dots & \frac{2}{N} & \frac{1}{N}\\
  T_{\text{dB}}^{i-}  & \frac{1}{N} & 0 & \dots & \frac{2}{N} & \dots & \frac{2}{N} & 0 \\
  t^{1 i}  & \frac{N}{3} & N\frac{N-3}{3} & \dots & N\frac{N-i-1}{3} & \dots & \frac{N}{3} & \frac{2N}{3}
 \end{array}\label{eq:Dominance_dB_07}
\end{eqnarray}
such that the mean fixation times become
\begin{eqnarray}
	t^1_\text{dB} &\to& \frac{N^3-5N^2+12N}{6} \label{eq:Dominance_dB_08}\\
	t^{1\to N}_\text{dB} &\to& \frac{3N^3-15N^2+34N}{12} \label{eq:Dominance_dB_09}
\end{eqnarray}
Hence for large population size, conditional fixation takes on average 50\% longer than unconditional fixation. 

Mutant-fixation with non-vanishing probability in the {\bf birth-death} process requires the four conditions $2b>c+d$, $a+b>2d$, $a<b$, and $a+b>2c$. As a result, we find the following transition probabilities and sojourn times 
\begin{eqnarray}
 \begin{array}{c|ccccccc}
    & 1 & 2 & \dots & i & \dots & N-2 & N-1  \\ \hline
  T_{\text{Bd}}^{i+} & 1 & \frac{1}{2} & \dots & \frac{1}{2} & \dots & \frac{1}{2} & \frac{1}{2}\\
  T_{\text{Bd}}^{i-}  & 0 & 0 & \dots & 0 & \dots & 0 & 0 \\
  t^{1 i}  & 1 & 2 & \dots & 2 & \dots & 2 & 2
 \end{array}\label{eq:Dominance_Bd_05}
\end{eqnarray}
The mutant fixes with probability one. 
We obtain fixation times linear in population size 
\begin{eqnarray}
	t^1_\text{Bd} = t^{1\to N}_\text{Bd}  \to 2N-3\label{eq:Dominance_Bd_06}
\end{eqnarray}
We can also find generic payoff matrices in which the mutant neither spreads nor goes extinct, despite strategic dominance of one strategy.

\subsection{Coordination games}

In coordination games, the interactions between two individuals of the same type always yield higher payoffs than interactions between two different types, hence the payoff relations for all coordination games are $a>c$ and $b<d$. 
In a well-mixed population, $A$ can then not invade $B$ and vice versa. 
Coordination games are used to study evolution of technological standards \citep{cooper:book:1998}, emerge in a subset of strategies in collective risk dilemmas \citep{milinski:PNAS:2008,hilbe:PlosOne:2013a}, and have a population genetic equivalent in genetic underdominance (heterozygote disadvantage) between two alleles of the same gene \citep{altrock:PLoSCB:2011}. 

In the {\bf death-birth} process, the limiting cases of the transition matrix (\ref{eq:dBStrong_01a})-(\ref{eq:dBStrong_01d}) immediately tell us that a mutant $A$ cannot invade, and we obtain
\begin{eqnarray}
 \begin{array}{c|ccccccc}
    & 1 & 2 & \dots & i & \dots & N-2 & N-1  \\ \hline
  T_{\text{dB}}^{i+} & 0 & 0 & \dots & 0 &0 & 0 & \frac{1}{N}\\
  T_{\text{dB}}^{i-}  & \frac{1}{N} & \frac{2}{N} & \dots & \frac{2}{N} & \dots & \frac{2}{N} & 0 \\
  t^{1 i}  & N & 0 & \dots & 0 & \dots & 0 & 0
 \end{array}\label{eq:Coordination_dB_01}
\end{eqnarray}
Thus, the conditional mean fixation time is not a meaningful quantity to compute. The unconditional mean fixation time, which only measures extinction of $A$, follows as 
\begin{eqnarray}
	t^1_\text{dB} \to N 
\label{eq:Coordination_dB_02}
\end{eqnarray}
This asymptotic relation holds for all coordination games under strong selection.
 
The {\bf birth-death} process leads to limiting cases in which the single mutant cannot invade. If the inequalities $c>d$ and $d+c>2b$ hold, we discover the following transition probabilities and sojourn times
\begin{eqnarray}
 \begin{array}{c|ccccccc}
    & 1 & 2 & \dots & i & \dots & N-2 & N-1  \\ \hline
  T_{\text{Bd}}^{i+} & 0 & \frac{1}{2} & \dots & 0 &0 & 0 & 0\\
  T_{\text{Bd}}^{i-}  & \frac{1}{2} & 0 & \dots & 0 & \dots & 0 & 0 \\
  t^{1 i}  & 2 & 0 & \dots & 0 & \dots & 0 & 0
 \end{array}\label{eq:Coordination_Bd_03}
\end{eqnarray}
The mutant goes extinct with probability one in two time steps on average,
\begin{eqnarray}
	t^1_\text{Bd} \to 2 
\label{eq:Coordination_Bd_02}
\end{eqnarray}

\subsection{Coexistence}

In coexistence games, it is always better for a individual to interact with an individual that plays a different strategy. The payoff relations are $a<c$ and $b>d$. Such strategic interactions emerge when cooperators generate a benefit that can be exploited by both cooperators and defectors \citep{maclean:PloSB:2010}. A coexistence game was the original motivation to study a game in an evolutionary context, namely the Hawk-Dove game \citep{maynard-smith:Nature:1973}, which allows a coexistence of such behaviors, in which common strategies outperform rare strategies \citep{doebeli:EL:2005}.  

In order to observe non-vanishing fixation probability of a single $A$ mutant in the {\bf death-birth} process, we require the following payoff-conditions to hold: $a+b > c+d$, $a+b>2d$, and $c+d>2a$ as an additional relation that determines the speed to fixation. Then, we can summarize the limiting cases of transition probabilities and sojourn times as follows
\begin{eqnarray}
 \begin{array}{c|ccccccc}
    & 1 & 2 & \dots & i & \dots & N-2 & N-1  \\ \hline
  T_{\text{dB}}^{i+} & \frac{2}{N} & \frac{2}{N} & \dots & \frac{2}{N} & \dots & \frac{2}{N} & \frac{1}{N}\\
  T_{\text{dB}}^{i-}  & \frac{1}{N} & 0 & \dots & \frac{2}{N} & \dots & \frac{2}{N} & \frac{2}{N} \\
  t^{1 i}  & \frac{N}{3} & N\frac{N-1}{3} & \dots & N\frac{N-i+1}{3} & \dots & N & \frac{2N}{3}
 \end{array}\label{eq:Coexistence_dB_05}
\end{eqnarray}
In this case of coexistence game, the mean unconditional fixation time sums up to a cubic polynomial in $N$
\begin{eqnarray}
	t^1_\text{dB} \to N^2\frac{N-1}{6} 
\label{eq:Coexistence_dB_06}
\end{eqnarray}
and the mean conditional fixation time amounts to 
\begin{eqnarray}
	t^{1\to N}_\text{dB} \to N\frac{3N(N-1)-2}{12}. 
\label{eq:Coexistence_dB_07}
\end{eqnarray}
If the coexistence game does not fulfill $c+d>2a$ we obtain
\begin{eqnarray}
 \begin{array}{c|ccccccc}
    & 1 & 2 & \dots & i & \dots & N-2 & N-1  \\ \hline
  T_{\text{dB}}^{i+} & \frac{2}{N} & \frac{2}{N} & \dots & \frac{2}{N} & \dots & \frac{2}{N} & \frac{1}{N}\\
  T_{\text{dB}}^{i-}  & \frac{1}{N}  & 0 & \dots & 0 & \dots & 0 & \frac{2}{N} \\
  t^{1 i}  & \frac{N}{3} & \frac{N}{3} & \dots & \frac{N}{3} & \dots & N & \frac{2N}{3}
 \end{array}\label{eq:Coexistence_dB_08}
\end{eqnarray}
which is slightly different than (\ref{eq:Dominance_dB_04}). Thus
\begin{eqnarray}
	t^1_\text{dB} &\to& N\frac{N+2}{3}
\label{eq:Coexistence_10}\\
	t^{1\to N}_\text{dB} &\to& N\frac{3N+5}{6}
\label{eq:Coexistence_dB_09}
\end{eqnarray}
There are two generic regimes of payoffs which lead to significantly different fixation times in coexistence games. In the first parameter regime, the mutation can go extinct even after it has spread to intermediate frequencies, and the fixation times scale in leading order with $N^3$. In the second regime the times scale with $N^2$, because in (\ref{eq:Coexistence_dB_05}), the process can go down from $i$ again, but not so in (\ref{eq:Coexistence_dB_08}), which make the process more efficient. 

\begin{figure}[b]
\centering
\includegraphics[width=0.9\columnwidth]{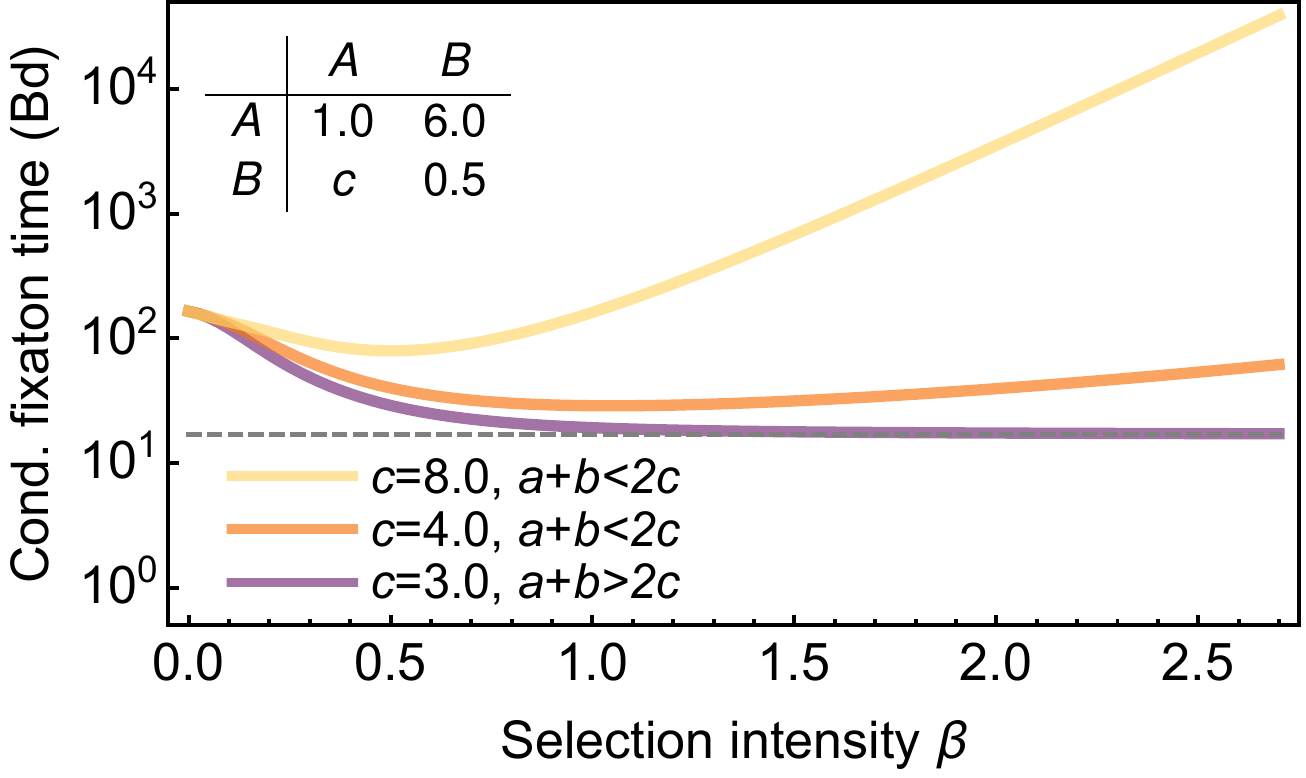}
\caption{\label{fig:Fig02}
(color online) \textbf{The conditional mean fixation time of the birth-death process in a coexistence game critically depends on payoff parameters}. If all conditions are met, reaction (\ref{eq:Dominance_Bd_06}) predicts the strong selection limit ($N=10$). However, the behavior with increasing selection strength can depend on a single parameter of the payoff matrix, e.g.~$c$, for which we show various curves $t_1^{1\to N}(\beta)$ (payoffs as inset). If $a+b>2c$ it follows that $t_{\text{Bd}}^{1\to N}\to 2\,N-3=17$ (dashed line), but if $a+b>2c$ it follows that $t_{\text{Bd}}^{1\to N}\to\infty$ ($\sim\text{e}^{\text{const.}\times\beta}$). The neutral mean fixation time is $N(N-1)/2$ and for intermediate selection strength, speedup can be observed \cite{altrock:JTB:2012}.
}
\end{figure}
The {\bf birth-death} process under a coexistence game requires the following five relations between payoffs in order to lead to non-vanishing fixation probability of the mutant: $2b>c+d$, $a+b>c+d$, $a+b>2d$, $a<b$, and $a+b>2c$, which together result in $\phi^1\to1$ and the same transition probabilities, sojourn times and fixation times we have already discovered for strategic dominance, Table (\ref{eq:Dominance_Bd_05}). Note that, if we relaxed the condition $a+b>2c$, such that $T_{\text{Bd}}^{(N-1)+}\to0$ and $T_{\text{Bd}}^{(N-1)-}\to1$, fixation would take infinitely long to occur. Similarly, in coexistence games with $a+b<c+d$, we find $T_{\text{Bd}}^{(i>1)+}\to0$ but $T_{\text{Bd}}^{1+}\to1$, and at the same time $T_{\text{Bd}}^{1-}\to0$ but $T_{\text{Bd}}^{(i>1)-}\to1/2$, $T_{\text{Bd}}^{(N-1)-}\to1$: the process gets trapped between the states 1 and 2, see Figure \ref{fig:Fig02}.

In coexistence games the intuition is that mean fixation times tend to infinity \cite{traulsen:PRE:2006c}. For both the death-birth and the birth death process on the cycle graph, we have shown that there are generic subsets of game parameters that allow fixation of the mutant with non-vanishing probability in a finite amount of time. Compared to a dominance game, the time scales in such coexistence games may be longer by a factor $N$ in the death-birth process, but identical for the birth-death process.

\subsection{Prisoner's dilemma vs. snowdrift game}

\begin{figure}[h]
\centering
\includegraphics[width=0.9\columnwidth]{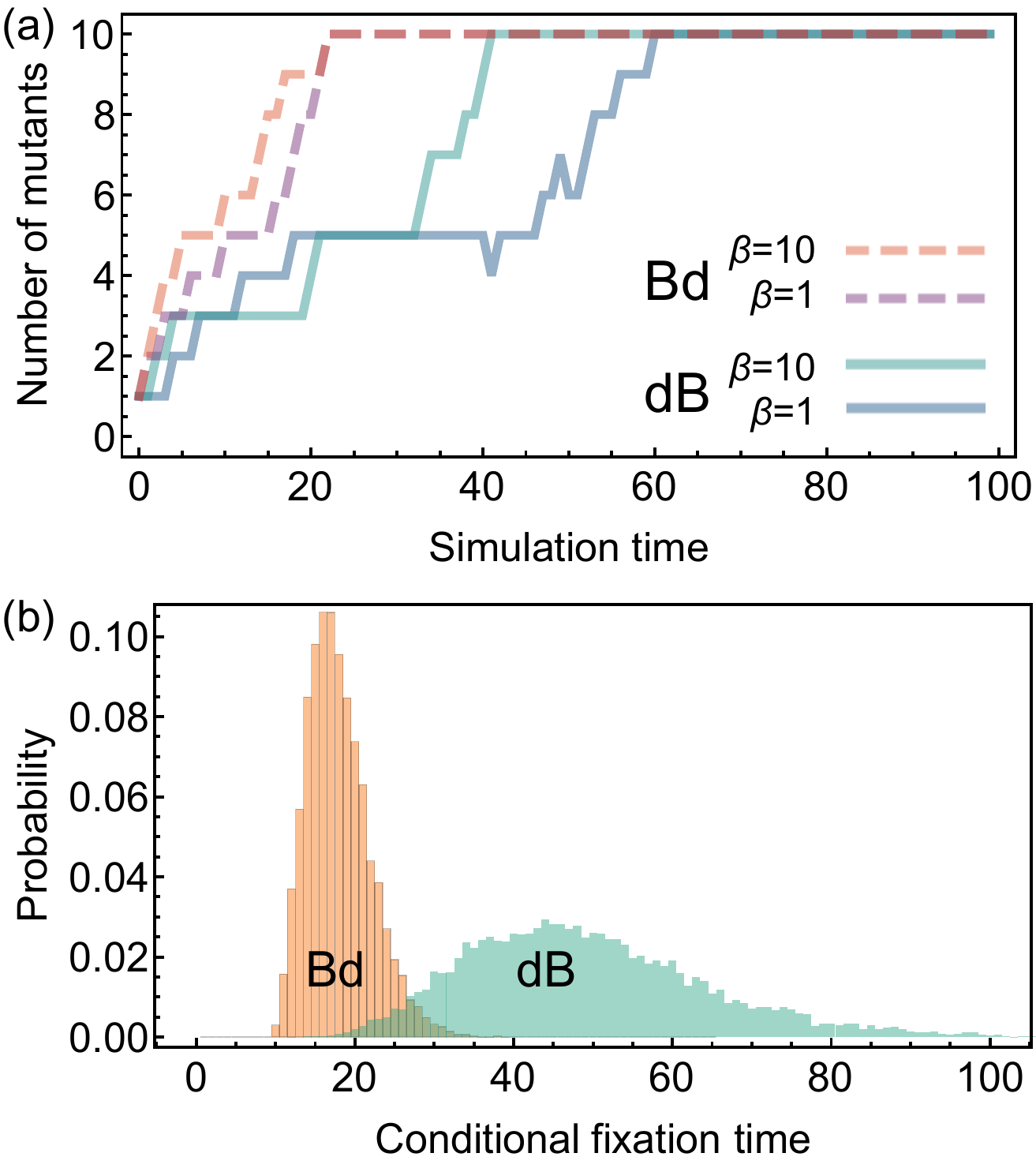}
\caption{\label{fig:Fig03}
(color online) \textbf{Trajectories and distributions of conditional fixation times in the Prisoner's dilemma game}, determined from simulations. The population size was $N=10$ individuals on a cycle. The payoff matrix in Eq.~(\ref{eq:payoffmatrixPD}) was used with benefit$=8$ and cost$=5$. a: Single independent realizations in which the defective mutation reached fixation for the dB process and the Bd process, darker shade $\beta=1$, lighter shade $\beta=10$. b: Histograms of $10^4$ independent realizations in which the defective mutation reached fixation. The mean conditional fixation time of the death-birth (dB) process is 48.4 (standard deviation 15.5), while for the birth-death (Bd) process it is 16.9 (standard deviation 4.0). In the dB case, (\ref{eq:Dominance_dB_04}) yields $t^{1\to N}\to N^2/2-N/6\approx48.3$, while in the Bd case, (\ref{eq:Dominance_Bd_05}) yields $t^{1\to N}\to 2\,N-3=17$ ($\beta=10$).
}
\end{figure}

We now turn to two concrete examples of social dilemma situations. We denote cooperators as the resident type and defectors as the mutant type. First, a game between a cooperative resident strategy $C$ and a defective mutant strategy $D$ is the prisoners' dilemma game, where cooperation corresponds to to offer a benefit to the co-player at a cost smaller than the benefit:
\begin{eqnarray}
 \begin{array}{c|cc}
  & D & C \\ \hline
  D & 0 & \text{benefit} \\
  C & -\text{cost} & \text{benefit-cost}
 \end{array}\label{eq:payoffmatrixPD}
\end{eqnarray}
The distribution of the conditional fixation times of this game, for both death-birth and birth-death update rules are shown in Figure \ref{fig:Fig03}. We assumed benefit$=8$ and cost$=5$ and measured the mean conditional fixation times for the death-birth and the birth-death processes. Due to the inherent stochasticity of the death-birth process even under very strong selection, the fixation time is considerably larger and more variable (compare (\ref{eq:Dominance_dB_04}) with (\ref{eq:Dominance_Bd_05})).

\begin{figure}[h!]
\centering
\includegraphics[width=0.9\columnwidth]{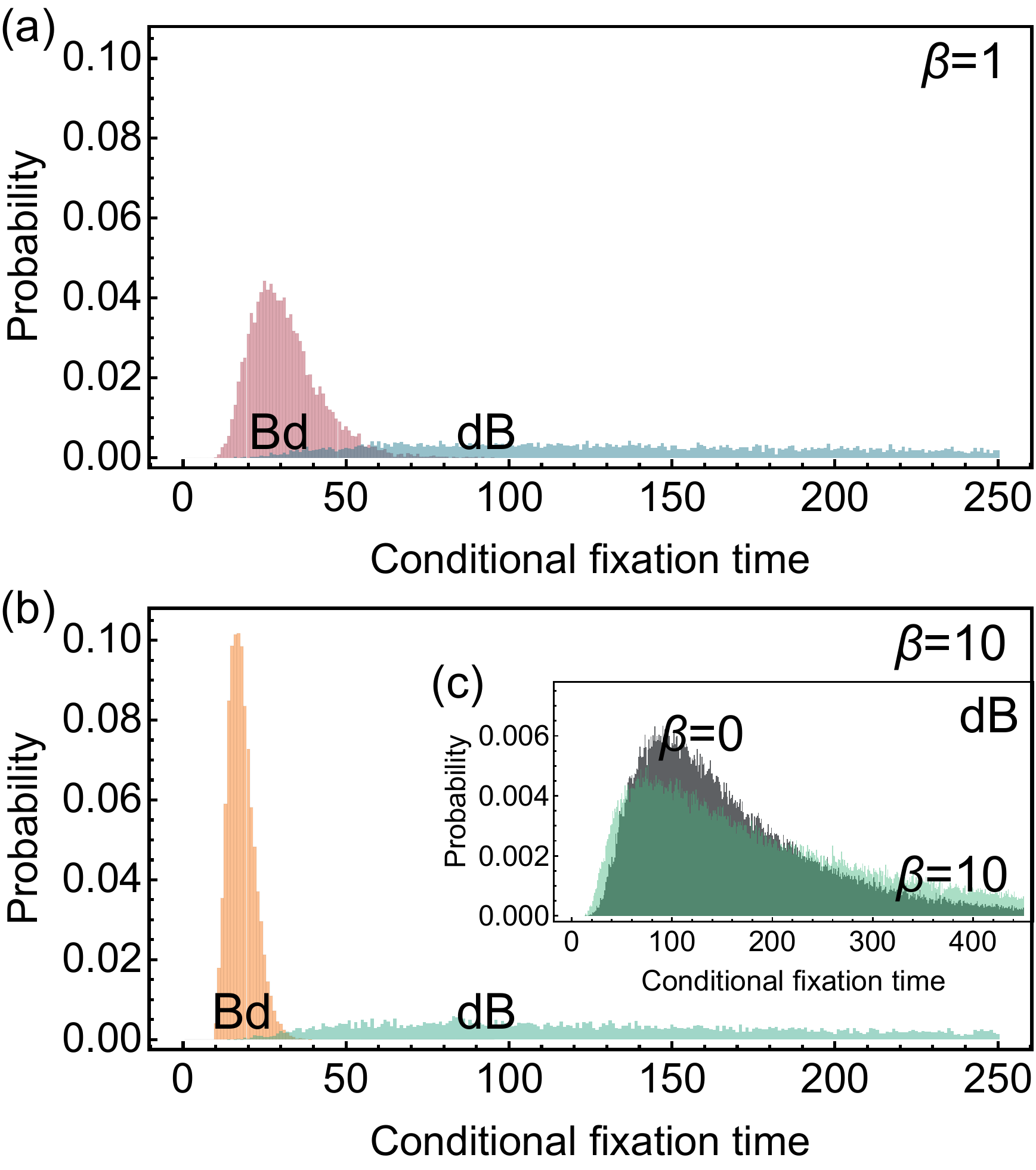}
\caption{\label{fig:Fig04}
(color online) \textbf{Conditional fixation time distributions in the snowdrift game}, obtained from simulations. The population size was $N=10$ individuals on the cycle. We show histograms of $10^4$ independent realizations in which the defector type reached fixation. The payoff matrix (\ref{eq:payoffmatrixSD}) was used. a: For $\beta=1$, the mean conditional fixation time of the death-birth (dB) process is 263.2 (standard deviation 198.7), while for the birth-death (Bd) process it is 10.3 (standard deviation 10.7). b: For $\beta=10$, the mean conditional fixation time of the death-birth (dB) process is 218.1 (standard deviation 168.9), while for the birth-death (Bd) process it is 17.0 (standard deviation 4.0). In the dB case, (\ref{eq:Dominance_dB_09}) yields $t^{1\to N}\approx223.3$, while in the Bd case, (\ref{eq:Dominance_Bd_06}) yields $t^{1\to N}\to 2\,N-3=17$. c: For neutral evolution, $\beta=0$, the mean conditional fixation time is 165 according to Eq.~(\ref{eq:neutral_03}) (only death-birth process shown). From the simulates we calculated a mean of 158.3 (standard deviation 92.0). These values are smaller than in the strong selection case, and the fixation time distribution is concentrated around smaller values and rare events of long fixation time are less common.
}
\end{figure}

Second, in the snowdrift game there is a benefit to the cooperator even when facing a defector. Cooperation still comes at a cost, but this cost is shared under mutual cooperation. A respective example payoff matrix of this form of coexistence game is
\begin{eqnarray}
 \begin{array}{c|cc}
  & D & C \\ \hline
  D & 1 & 8 \\
  C & 3 & 4
 \end{array}\label{eq:payoffmatrixSD}
\end{eqnarray}
This payoff configuration leads to the cases (\ref{eq:Coexistence_dB_05}) for the death-birth process, and to (\ref{eq:Dominance_Bd_05}) for the birth-death process. Hence with Eq.~(\ref{eq:Coexistence_dB_06}) we expect long fixation times in the death-birth precess, and fast fixation in the birth-death process, Eq.~(\ref{eq:Dominance_Bd_06}). Our simulations show that fixation times in the death-birth process of this coexistence game tend to be extremely widely distributed, Figure \ref{fig:Fig04}, whereas the death-birth processes fixation times are narrowly distributed around the mean for strong selection. For comparison, in the neutral case, \ref{fig:Fig04} (c), rare events of long fixation times are less common.

These examples show that while in birth-death driven evolutionary dynamics in structured populations an advantageous mutant can take over quickly under strong selection, the inherent stochasticity of the death-birth driven dynamics can lead to a wide distribution of fixation times. For coexistence games, birth-death processes can again lead to fast fixation in times of order $N$, and death-birth process fixes in times of order $N^3$, but with a very wide distribution. This leading order is much smaller than in the well-mixed population case, in which fixation times tend to infinity for strong selection \cite{antal:BMB:2006,altrock:JTB:2012}.

\section{\label{sec:conclusion}Summary and Conclusions}

Selective forces of stochastic evolutionary dynamics in structured populations are driven by the underlying update rule \cite{ohtsuki:Nature:2006,frean:PRSB:2013,hindersin:PlosCB:2015} and by how payoff is translated into fitness \cite{maciejewski:PLoSCB:2014}. Here we focus on strong selection on the cycle graph \cite{ohtsuki:PRSB:2006} and an exponential payoff to fitness mapping. Using both analytical calculations and simulations, we show that outcomes of the death-birth process may differ drastically from outcomes of the birth-death process. Under strong selection, transitions in the death-birth process remain stochastic (proportional to $1/N$), as the random death-step before selection always depends on population size. The maintenance of this degree of demographic noise is surprising, as for neutral evolution conditional fixation times are of leading order $N^3$, and there are games for which the strong selection fixation times are also of this leading order (or of leading order $N^2$). Under the birth-death process, fitness effects can eliminate dependence on population size, and transition rates become constant under strong selection. Hence, the leading order of mean fixation times is $N$. 
\begin{figure}[h!]
\centering
\includegraphics[width=0.95\columnwidth]{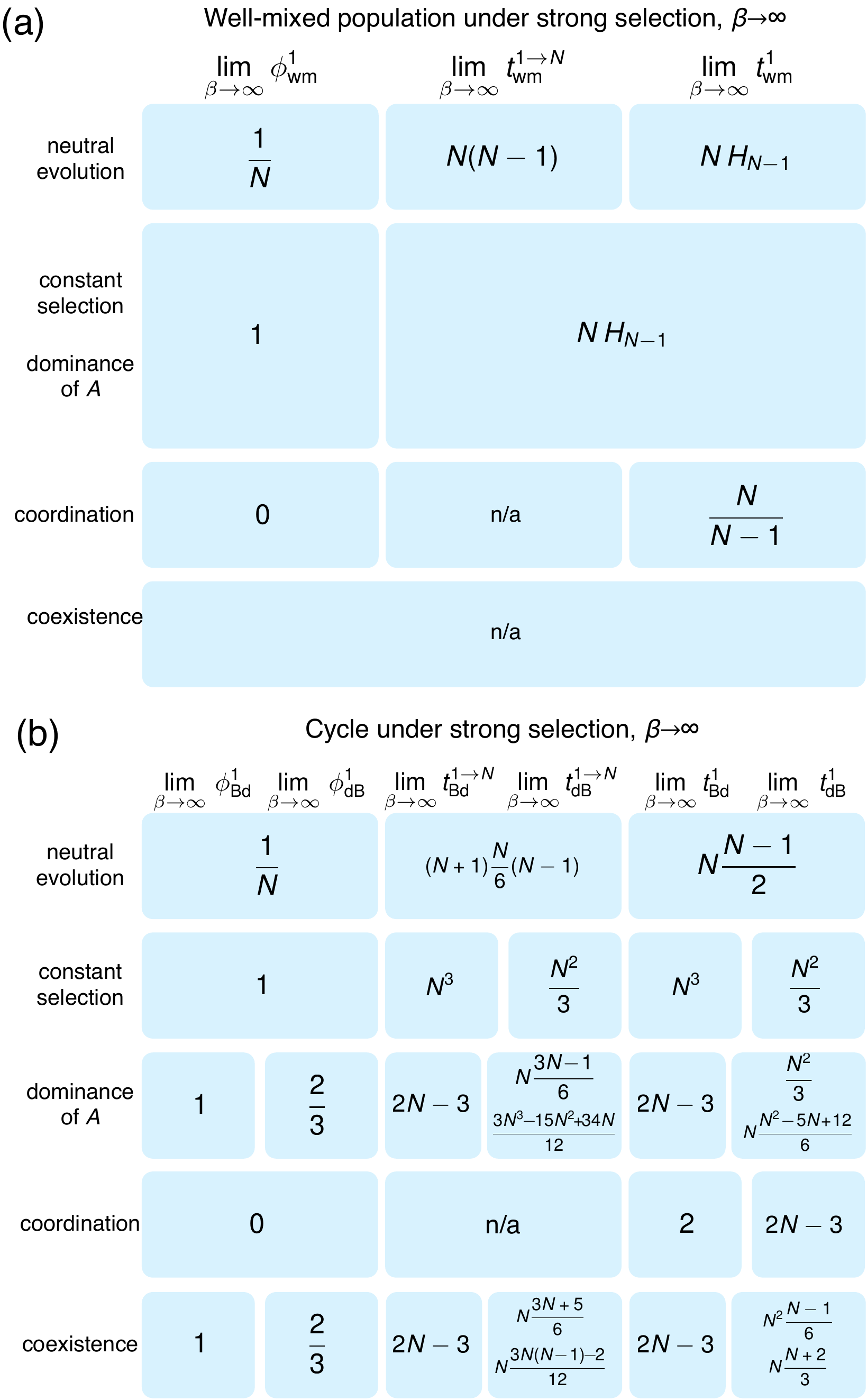}
\caption{\label{fig:Fig05}
(color online) \textbf{Overview of fixation probability and times in the strong selection limit, comparing the well-mixed population with dynamics on the cycle.} 
a: In case of the well-mixed population, the results have been worked out previously \cite{altrock:JTB:2012}. Most importantly, for coexistence games the fixation times under strong selection tend to infinity for any population size. 
b: In spatial populations, the update process matters. 
Already the constant selection cases show differences comparing Bd to dB update processes. 
For strong frequency dependent selection not only fixation times, but also fixation probabilities can differ. 
As opposed to the well-mixed population, there are generic subsets of coexistence games for which fixation occurs in a finite number of time steps, with averages depending on the game. This leads to distinct strong selection limit cases (see \ref{sec:discussion}). E.g., compare Eq.~(\ref{eq:Coexistence_dB_07}) (conditional fixation time is of leading order $N^3$) with Eq.~(\ref{eq:Coexistence_dB_09}) (conditional fixation time is of leading order $N^2$). Note that there also are coexistence games with payoff configurations such that strong selection works entirely against fixation of the mutant (comparable to coordination games).
}
\end{figure}
For dominance games, both the death-birth and the birth death process can lead to non-vanishing fixation probability and finite fixation times when the payoff of mutants interacting with a resident, $b$, is sufficiently large compared to all other payoffs. This condition ensures that the mutant can invade and stays advantageous at the boundary between residents and mutants. For the example of a strategic dominance game, such as the the prisoner's dilemma, not all payoff configurations allow mutant invasion \cite{ohtsuki:PRSB:2006}: if the cost of cooperation is sufficiently low, a population of cooperators is immune to invasion and fixation of a defective mutant under both update processes. This result stands in contrast to the well-mixed population. If the payoff configuration is such that defective mutants can invade, i.e for high benefit and low cost, the speed of fixation is expected to take long and fixation remains highly stochastic for the death-birth process, even under strong selection. 
For a birth-death update driven process, rapid takeover by an advantageous mutant can be observed, see Figure \ref{fig:Fig05}.

Stochastic evolutionary dynamics of coordination games on the cycle graph show closest resemblance to their well-mixed counterpart: under strong selection, mutants cannot invade, and the respective extinction times are short (of order $N$ in the death-birth process, of order 1 in the birth-death process), see Figure \ref{fig:Fig05}. The snowdrift game promotes coexistence of cooperators and defectors in well-mixed populations under strong selection \cite{altrock:PRE:2010,altrock:JTB:2012}. For this game on the cycle the evolutionary dynamics can get trapped at intermediate numbers of mutants of either kind and fixation would also take infinitely long. There are coexistence game payoff configurations that  permit mutant invasion, on the cycle, again when $b$ is the dominating payoff. In such cases, fixation times of the death-birth process tend to be much longer and experience a broader distribution than fixation times of the birth-death process. 

Our analytical results focus on mean fixation times, but the simulation results highlight that future work has to focus on other features of the fixation time distribution, such as higher moments. Overall our work highlights that selective advantage might not guarantee fixation within a desired time frame, even on graphs that are deemed not to be a suppressor or amplifier of selection.

\section*{Acknowledgements}
We thank Christian Hilbe, Kamran Kaveh and two anonymous referees for helpful comments.

P.M.A.~gratefully acknowledges financial support from Deutsche Akademie der Naturforscher Leopoldina, grant no.~LPDS 2012-12. 
A.T.~is grateful for generous funding by the Max-Planck Society. 
M.A.N.~was supported by the Bill and Melinda Gates Foundation [OPP1148627], the Office of Naval Research grant N00014-16-1-2914, National Cancer Institute grant CA179991 and by the John Templeton Foundation. The Program for Evolutionary Dynamics is supported in part by a gift from B.~Wu and Eric Larson.

\newpage
\begin{widetext}
\appendix

\section{\label{sec:Bdtransitions}Transition probabilities of the birth-death process}

The transition probabilities of the birth-death process on the cycle graph for any strength of selection, in full form, are given by
\begin{eqnarray}\label{eq:BdRing_p}
\begin{split}
T_{\text{Bd}}^{i+}= 
\begin{cases}
\frac{\text{e}^{\beta\,2 b}}{\text{e}^{ \beta\,2 b}+2 \text{e}^{\beta(c+d)}+(N-3) \text{e}^{\beta\,2d}}		\,\,\,&\text{if}\,\,\,i=1\\
\frac{\text{e}^{\beta  (a+b)}}{2 \text{e}^{\beta  (a+b)}+(i-2) \text{e}^{\beta\,2 a  }+2 \text{e}^{\beta  (c+d)}+(N-i-2)\text{e}^{\beta \,2 d} }		\,\,\,&\text{if}\,\,\,1<i<N-1\\
\frac{\text{e}^{\beta  (a+b)}}{2 \text{e}^{\beta  (a+b)}+(N-3) \text{e}^{\beta\,2 a  }+\text{e}^{\beta\,2 c}}		\,\,\,&\text{if}\,\,\,i=N-1\\
0\,\,\,&\text{if}\,\,\,i=0,N
\end{cases}
\end{split}
\end{eqnarray}
and 
\begin{eqnarray}\label{eq:BdRing_m}
\begin{split}
T_{\text{Bd}}^{i-} = 
\begin{cases}
\frac{\text{e}^{\beta  (c+d)}}{\text{e}^{\beta\,2 b}+2 \text{e}^{\beta  (c+d)}+(N-3) \text{e}^{\beta\,2 d}}		\,\,\,&\text{if}\,\,\,i=1\\
\frac{\text{e}^{\beta  (c+d)}}{2 \text{e}^{\beta  (a+b)}+(i-2) \text{e}^{\beta\,2 a  }+2 \text{e}^{\beta  (c+d)}+(N-i-2)\text{e}^{\beta\,2 d} }		\,\,\,&\text{if}\,\,\,1<i<N-1\\
\frac{\text{e}^{\beta\,2 c}}{2 \text{e}^{\beta  (a+b)}+(N-3) \text{e}^{\beta\,2 a  }+\text{e}^{\beta\,2 c}}		\,\,\,&\text{if}\,\,\,i=N-1\\
0		\,\,\,&\text{if}\,\,\,i=0,N
\end{cases}
\end{split}
\end{eqnarray}

\section{\label{sec:probabilities}The probability to ever visit $j$, starting from $i$}

The probabilities to ever go from any internal state $i$ to any other internal state $j$ are useful when calculating sojourn times \cite{ewens:book:2004,altrock:JTB:2012},
 \begin{eqnarray}
	\phi^{i j}&=&\frac{
	\sum _{k=i}^{N-1} \prod _{m=j+1}^k \frac{T^{m-}}{T^{m+}}
	}{
	\sum _{k=j}^{N-1} \prod_{m=j+1}^k \frac{T^{m-}}{T^{m+}}
	}\,\,\,\text{if}\,\,\,i>j, \label{eq:sojourn_05}\\
	\phi^{i j}&=&\frac{
	\sum _{k=0}^{i-1} \prod _{m=1}^k \frac{T^{m-}}{T^{m+}}
	}{
	\sum _{k=0}^{j-1} \prod_{m=1}^k \frac{T^{m-}}{T^{m+}}
	}\,\,\,\text{if}\,\,\,i<j, \label{eq:sojourn_06}
\end{eqnarray}
which hold for any birth-death process with absorbing boundaries \citep{goel:book:1974}. 

\section{\label{sec:constant}Constant selection on cycle graphs}

We speak of constant selection if the fitness of mutants is $f_A>1$ and the fitness of residents is $f_B\equiv1$, irrespective of the number of mutants. Often, the success of the mutant is given by $r>1$, where the fitness of the resident is 1. We can then define $q=f_B/f_A=\text{e}^{-\beta r}$, and strong constant selection ($\beta\,r\to\infty$) in favor go the mutant means $q\to0$. In the following we derive exact expressions for transition probabilities, sojourn times and the fixation time under strong constant selection.

In case of the {\bf birth-death} process we find $q=T_{\text{Bd}}^{j-}/T_{\text{Bd}}^{j+}$ for all $j$ and get $\phi^{i+1,i}_\text{Bd}=1-(1-q)/(1-q^{N-i})$, and $\phi^{i,i+1}_\text{Bd}=(1-q^i)/(1-q^{i+1})$. Then, we obtain the fixation probability \cite{frean:PRSB:2013,hindersin:PlosCB:2015,askari:PRE:2015}
\begin{eqnarray}\label{eq:constant01x}
\phi^i_\text{Bd} = \frac{1-q^i}{1-q^N},
\end{eqnarray}
as well as the sojourn times
\begin{eqnarray}\label{eq:constant02x}
t^{1j}_\text{Bd} = \frac{1}{T^{j+}_{\text{Bd}}(q)}\frac{1-q}{1-q^j}\frac{1}{
\frac{1-q}{1-q^{N-j}}+q\left( 1-\frac{1-q^{j-1}}{1-q^j} \right)
}
\end{eqnarray}
In the strong selection limit $q\to0$ we obtain $T^{j+}_{\text{Bd}}\to1/j$, and we can immediately see that 
\begin{eqnarray}\label{eq:constant03}
t^{1j}_\text{Bd} \to j
\end{eqnarray}
and summing over all sojourn times case leads to
\begin{eqnarray}\label{eq:constant04}
t^{1}_\text{Bd} \to N\frac{N-1}{2}
\end{eqnarray}
which is equal to $t^{1\to N}_\text{Bd}$ as the mutant fixes with probability 1.

For the {\bf death-birth process}, we obtain $T^{1-}_{\text{dB}}/T^{1+}_{\text{dB}}=(1+q)/2$, $T^{i-}_{\text{dB}}/T^{i+}_{\text{dB}}=q$ for $1<i<N-1$, and $T^{(N-1)-}_{\text{dB}}/T^{(N-1)+}_{\text{dB}}=2q/(1+q)$. Since the first and last state are special due to the structure of random death and subsequent competition, we obtain a more complicated expression for the fixation probability
\begin{eqnarray}\label{eq:constant05}
\phi^i_\text{dB} = 
\frac{
1+\frac{1+q}{1-q}\frac{1-q^{i-1}}{2}
}{
1+q^{N-2}+\frac{1+q }{1-q}\frac{1-q^{N-2}}{2}
}
\end{eqnarray}
With the appropriate values of $\phi^{j+1,j}_\text{dB}$ and $\phi^{j,j+1}_\text{dB}$, the expected sojourn times become 
\begin{eqnarray}
t^{1\,1}_\text{dB} &=&\frac{1}{
\frac{2}{N}\frac{1}{1+q}\left( 1-\phi^{2,1}_\text{dB} \right)+\frac{1}{N}
}\label{eq:constant06a}\\
t^{1\,j}_\text{dB} &=& 
\frac{2 (1-q)}{3-q-(1+q) q^{j-1}}
\frac{1}{
\frac{2}{N}\frac{1}{1+q}\left( 1-\phi^{j+1,j}_\text{dB} \right)+\frac{2}{N}\frac{q}{1+q}\left( 1-\phi^{j-1,j}_\text{dB} \right)
}\label{eq:constant06b}\\
t^{1\,(N-1)}_\text{dB} &=& 
\frac{2 (1-q)}{3-q-(1+q) q^{N-2}}
\frac{1}{
\frac{1}{N}+\frac{2}{N}\frac{q}{1+q}\left( 1-\phi^{N-2,N-1}_\text{dB} \right)
}\label{eq:constant06c}
\end{eqnarray}
For strong selection, $q\to0$, we observe that $\phi^i_\text{dB} \to1$, $\phi^{j+1,j}_\text{dB} \to0$ and $\phi^{j-1,j}_\text{dB} \to\text{const.}$ for all $j$, respectively. For the sojourn times we can now see that 
\begin{eqnarray}
t^{1\,1}_\text{dB} &\to& \frac{N}{3}  \label{eq:constant07a}\\
t^{1\,j}_\text{dB} &\to& \frac{N}{3}  \label{eq:constant07b}\\
t^{1\,(N-1)}_\text{dB} &\to& \frac{2\,N}{3}  \label{eq:constant07c}\\
\end{eqnarray}
and hence the mean fixation time amount to
\begin{eqnarray}\label{eq:constant08}
t^{1\to N}_\text{dB}=t^{1}_\text{dB} \to\frac{N^2}{3}
\end{eqnarray}

\section{\label{sec:constantExpansion}Constant selection approximation for the Bd process (cycle and well-mixed)}

Here we are interested in finding approximations of the conditional mean fixation times for finite but large values of the product $\beta\,r$. In this case, we can perform a Taylor expansion to linear order in the quantity $q=\text{e}^{-\beta\,r}=f_B/f_A$, as defined above. For the {\bf well-mixed} case we obtain $\phi^1_\text{WM}\approx1-q$ and $\phi^{i>1}_\text{WM}\approx1$, as well as $T_{\text{WM}}^{i-}/T_{\text{WM}}^{i+}=q$ for all $0<i<N$. To calculate the conditional mean fixation time of the well-mixed population we here resort to an equation that does not involve mean sojourn times \cite{goel:book:1974}
\begin{eqnarray}\label{eq:constant09}
t^{1\to N}_\text{WM} = \sum\limits_{k=1}^{N-1} \sum\limits_{l=1}^{k}\frac{\phi_{\text{WM}}^l}{T_{\text{WM}}^{l+}}\,q^{k-j}
\end{eqnarray}
and sort this equation in all terms involving directly $q^0$ or $q^1$
\begin{eqnarray}\label{eq:constant10}
t^{1\to N}_\text{WM} = \sum\limits_{k=1}^{N-1}s_k(q) + q\,\sum\limits_{k=1}^{N-2}s_k(q)
\end{eqnarray}
whereby $s_k(q)=\phi_{\text{WM}}^k/T_{\text{WM}}^{k+}$. Now we use Eq.~(\ref{eq:Moran_01a}) with $\text{e}^{\beta\pi_A}=\text{e}^{\beta\,r}$ and $\text{e}^{\beta\pi_B}=1$, which is equivalent to setting the payoff $a=b=r/2$ and $c=d=0$, we get the linear in $q$ approximation
\begin{eqnarray}\label{eq:constant11}
\frac{1}{T_{\text{WM}}^{k+}} \approx \frac{N}{N-k}+\frac{N}{k}q
\end{eqnarray}
to result in 
\begin{eqnarray}
s_1\ \approx \frac{N}{N-1}-q\frac{N}{N-1}+q\,N \label{eq:constant12a}\\
s_{k>1}\ \approx \frac{N}{N-k} + \frac{N}{k}q \label{eq:constant12a}
\end{eqnarray}
Then, the first sum of Eq.~(\ref{eq:constant10}) amounts to 
\begin{eqnarray}\label{eq:constant12}
\sum\limits_{k=1}^{N-1}s_k(q) \approx N\,H_{N-1}+q\,N\,H_{N-2}
\end{eqnarray}
and the second sum solves similarly. We thus obtain 
\begin{eqnarray}\label{eq:constant13}
t^{1\to N}_\text{WM} \approx N\,H_{N-1}+2\,q\,N\,H_{N-2}
\end{eqnarray}
In case of the {\bf birth-death} process on the cycle, we use the mean sojourn times from Eq.~(\ref{eq:constant02x}), the definition of the mean fixation time of Eq.~(\ref{eq:sojourn_04}), and $1/T_{\text{Bd}}^{k+}=k+(N-k)q$, to result in 
\begin{eqnarray}\label{eq:constant14}
t^{1\to N}_\text{Bd} \approx N\frac{N-1}{2}+q\,N(N-2)
\end{eqnarray}

\end{widetext}


\begin{thebibliography}{51}
\expandafter\ifx\csname natexlab\endcsname\relax\def\natexlab#1{#1}\fi
\expandafter\ifx\csname bibnamefont\endcsname\relax
  \def\bibnamefont#1{#1}\fi
\expandafter\ifx\csname bibfnamefont\endcsname\relax
  \def\bibfnamefont#1{#1}\fi
\expandafter\ifx\csname citenamefont\endcsname\relax
  \def\citenamefont#1{#1}\fi
\expandafter\ifx\csname url\endcsname\relax
  \def\url#1{\texttt{#1}}\fi
\expandafter\ifx\csname urlprefix\endcsname\relax\def\urlprefix{URL }\fi
\providecommand{\bibinfo}[2]{#2}
\providecommand{\eprint}[2][]{\url{#2}}

\bibitem[{\citenamefont{Maynard~Smith and
  Price}(1973)}]{maynard-smith:Nature:1973}
\bibinfo{author}{\bibfnamefont{J.}~\bibnamefont{Maynard~Smith}}
  \bibnamefont{and} \bibinfo{author}{\bibfnamefont{G.~R.} \bibnamefont{Price}},
  \bibinfo{journal}{Nature} \textbf{\bibinfo{volume}{246}}, \bibinfo{pages}{15}
  (\bibinfo{year}{1973}).

\bibitem[{\citenamefont{Hofbauer and Sigmund}(1988)}]{hofbauer:book:1988}
\bibinfo{author}{\bibfnamefont{J.}~\bibnamefont{Hofbauer}} \bibnamefont{and}
  \bibinfo{author}{\bibfnamefont{K.}~\bibnamefont{Sigmund}},
  \emph{\bibinfo{title}{The theory of evolution and dynamical systems:
  {M}athematical aspects of selection}} (\bibinfo{publisher}{Cambridge
  University Press}, \bibinfo{address}{Cambridge}, \bibinfo{year}{1988}).

\bibitem[{\citenamefont{Taylor and Jonker}(1978)}]{taylor:MB:1978}
\bibinfo{author}{\bibfnamefont{P.~D.} \bibnamefont{Taylor}} \bibnamefont{and}
  \bibinfo{author}{\bibfnamefont{L.}~\bibnamefont{Jonker}},
  \bibinfo{journal}{Mathematical Biosciences} \textbf{\bibinfo{volume}{40}},
  \bibinfo{pages}{145} (\bibinfo{year}{1978}).

\bibitem[{\citenamefont{Hofbauer et~al.}(1979)\citenamefont{Hofbauer, Schuster,
  and Sigmund}}]{hofbauer:JTB:1979}
\bibinfo{author}{\bibfnamefont{J.}~\bibnamefont{Hofbauer}},
  \bibinfo{author}{\bibfnamefont{P.}~\bibnamefont{Schuster}}, \bibnamefont{and}
  \bibinfo{author}{\bibfnamefont{K.}~\bibnamefont{Sigmund}},
  \bibinfo{journal}{Journal of Theoretical Biology}
  \textbf{\bibinfo{volume}{81}}, \bibinfo{pages}{609} (\bibinfo{year}{1979}).

\bibitem[{\citenamefont{Zeeman}(1980)}]{zeeman:LN:1980}
\bibinfo{author}{\bibfnamefont{E.~C.} \bibnamefont{Zeeman}},
  \bibinfo{journal}{Lecture Notes in Mathematics}
  \textbf{\bibinfo{volume}{819}}, \bibinfo{pages}{471} (\bibinfo{year}{1980}).

\bibitem[{\citenamefont{Ohtsuki and
  Nowak}(2006{\natexlab{a}})}]{ohtsuki:JTB:2006b}
\bibinfo{author}{\bibfnamefont{H.}~\bibnamefont{Ohtsuki}} \bibnamefont{and}
  \bibinfo{author}{\bibfnamefont{M.~A.} \bibnamefont{Nowak}},
  \bibinfo{journal}{Journal of Theoretical Biology}
  \textbf{\bibinfo{volume}{243}}, \bibinfo{pages}{86}
  (\bibinfo{year}{2006}{\natexlab{a}}).

\bibitem[{\citenamefont{Ohtsuki and Nowak}(2008)}]{ohtsuki:JTB:2008}
\bibinfo{author}{\bibfnamefont{H.}~\bibnamefont{Ohtsuki}} \bibnamefont{and}
  \bibinfo{author}{\bibfnamefont{M.~A.} \bibnamefont{Nowak}},
  \bibinfo{journal}{Journal of Theoretical Biology}
  \textbf{\bibinfo{volume}{251}}, \bibinfo{pages}{698} (\bibinfo{year}{2008}).

\bibitem[{\citenamefont{Fogel et~al.}(1998)\citenamefont{Fogel, Andrews, and
  Fogel}}]{fogel:EM:1998}
\bibinfo{author}{\bibfnamefont{G.}~\bibnamefont{Fogel}},
  \bibinfo{author}{\bibfnamefont{P.}~\bibnamefont{Andrews}}, \bibnamefont{and}
  \bibinfo{author}{\bibfnamefont{D.}~\bibnamefont{Fogel}},
  \bibinfo{journal}{Ecological Modelling} \textbf{\bibinfo{volume}{109}},
  \bibinfo{pages}{283} (\bibinfo{year}{1998}).

\bibitem[{\citenamefont{Nowak et~al.}(2004)\citenamefont{Nowak, Sasaki, Taylor,
  and Fudenberg}}]{nowak:Nature:2004}
\bibinfo{author}{\bibfnamefont{M.~A.} \bibnamefont{Nowak}},
  \bibinfo{author}{\bibfnamefont{A.}~\bibnamefont{Sasaki}},
  \bibinfo{author}{\bibfnamefont{C.}~\bibnamefont{Taylor}}, \bibnamefont{and}
  \bibinfo{author}{\bibfnamefont{D.}~\bibnamefont{Fudenberg}},
  \bibinfo{journal}{Nature} \textbf{\bibinfo{volume}{428}},
  \bibinfo{pages}{646} (\bibinfo{year}{2004}).

\bibitem[{\citenamefont{Traulsen et~al.}(2005)\citenamefont{Traulsen, Claussen,
  and Hauert}}]{traulsen:PRL:2005}
\bibinfo{author}{\bibfnamefont{A.}~\bibnamefont{Traulsen}},
  \bibinfo{author}{\bibfnamefont{J.~C.} \bibnamefont{Claussen}},
  \bibnamefont{and} \bibinfo{author}{\bibfnamefont{C.}~\bibnamefont{Hauert}},
  \bibinfo{journal}{Physical Review Letters} \textbf{\bibinfo{volume}{95}},
  \bibinfo{pages}{238701} (\bibinfo{year}{2005}).

\bibitem[{\citenamefont{Hilbe}(2011)}]{hilbe:BMB:2011}
\bibinfo{author}{\bibfnamefont{C.}~\bibnamefont{Hilbe}},
  \bibinfo{journal}{Bulletin of Mathematical Biology}
  \textbf{\bibinfo{volume}{73}}, \bibinfo{pages}{2068} (\bibinfo{year}{2011}).

\bibitem[{\citenamefont{Black and McKane}(2012)}]{black:TREE:2012}
\bibinfo{author}{\bibfnamefont{A.~J.} \bibnamefont{Black}} \bibnamefont{and}
  \bibinfo{author}{\bibfnamefont{A.~J.} \bibnamefont{McKane}},
  \bibinfo{journal}{Trends in Ecology and Evolution}
  \textbf{\bibinfo{volume}{27}}, \bibinfo{pages}{337} (\bibinfo{year}{2012}).

\bibitem[{\citenamefont{Altrock et~al.}(2010)\citenamefont{Altrock, Gokhale,
  and Traulsen}}]{altrock:PRE:2010}
\bibinfo{author}{\bibfnamefont{P.~M.} \bibnamefont{Altrock}},
  \bibinfo{author}{\bibfnamefont{C.~S.} \bibnamefont{Gokhale}},
  \bibnamefont{and} \bibinfo{author}{\bibfnamefont{A.}~\bibnamefont{Traulsen}},
  \bibinfo{journal}{Physical Review E} \textbf{\bibinfo{volume}{82}},
  \bibinfo{pages}{011925} (\bibinfo{year}{2010}).

\bibitem[{\citenamefont{Lieberman et~al.}(2005)\citenamefont{Lieberman, Hauert,
  and Nowak}}]{lieberman:Nature:2005}
\bibinfo{author}{\bibfnamefont{E.}~\bibnamefont{Lieberman}},
  \bibinfo{author}{\bibfnamefont{C.}~\bibnamefont{Hauert}}, \bibnamefont{and}
  \bibinfo{author}{\bibfnamefont{M.~A.} \bibnamefont{Nowak}},
  \bibinfo{journal}{Nature} \textbf{\bibinfo{volume}{433}},
  \bibinfo{pages}{312} (\bibinfo{year}{2005}).

\bibitem[{\citenamefont{Broom and Rycht{\'a}\v{r}}(2008)}]{broom:PRSA:2008}
\bibinfo{author}{\bibfnamefont{M.}~\bibnamefont{Broom}} \bibnamefont{and}
  \bibinfo{author}{\bibfnamefont{J.}~\bibnamefont{Rycht{\'a}\v{r}}},
  \bibinfo{journal}{Proceedings of the Royal Society A}
  \textbf{\bibinfo{volume}{464}}, \bibinfo{pages}{2609} (\bibinfo{year}{2008}).

\bibitem[{\citenamefont{Broom et~al.}(2010)\citenamefont{Broom,
  Hadjichrysathou, and Rycht{\'a}\v{r}}}]{broom:PRSA:2010}
\bibinfo{author}{\bibfnamefont{M.}~\bibnamefont{Broom}},
  \bibinfo{author}{\bibfnamefont{C.}~\bibnamefont{Hadjichrysathou}},
  \bibnamefont{and}
  \bibinfo{author}{\bibfnamefont{J.}~\bibnamefont{Rycht{\'a}\v{r}}},
  \bibinfo{journal}{Proceedings of the Royal Society A}
  \textbf{\bibinfo{volume}{466}}, \bibinfo{pages}{1327} (\bibinfo{year}{2010}).

\bibitem[{\citenamefont{Hindersin and Traulsen}(2014)}]{hindersin:JRSI:2014}
\bibinfo{author}{\bibfnamefont{L.}~\bibnamefont{Hindersin}} \bibnamefont{and}
  \bibinfo{author}{\bibfnamefont{A.}~\bibnamefont{Traulsen}},
  \bibinfo{journal}{Journal of the Royal Society Interface}
  \textbf{\bibinfo{volume}{11}}, \bibinfo{pages}{20140606}
  (\bibinfo{year}{2014}).

\bibitem[{\citenamefont{Kaveh et~al.}(2015)\citenamefont{Kaveh, N.L.Komarova,
  and Kohandel}}]{kaveh:JRSOS:2015}
\bibinfo{author}{\bibfnamefont{K.}~\bibnamefont{Kaveh}},
  \bibinfo{author}{\bibnamefont{N.L.Komarova}}, \bibnamefont{and}
  \bibinfo{author}{\bibfnamefont{M.}~\bibnamefont{Kohandel}},
  \bibinfo{journal}{Journal of the Royal Society Open Science}
  \textbf{\bibinfo{volume}{2}}, \bibinfo{pages}{140465} (\bibinfo{year}{2015}).

\bibitem[{\citenamefont{Nowak}(2006)}]{nowak:book:2006}
\bibinfo{author}{\bibfnamefont{M.~A.} \bibnamefont{Nowak}},
  \emph{\bibinfo{title}{Evolutionary Dynamics}} (\bibinfo{publisher}{Harvard
  University Press}, \bibinfo{address}{Cambridge MA}, \bibinfo{year}{2006}).

\bibitem[{\citenamefont{Maruyama}(1970)}]{maruyama:GR:1970}
\bibinfo{author}{\bibfnamefont{T.}~\bibnamefont{Maruyama}},
  \bibinfo{journal}{Genetics Research} \textbf{\bibinfo{volume}{15}},
  \bibinfo{pages}{221} (\bibinfo{year}{1970}).

\bibitem[{\citenamefont{Nei and Roychoudhury}(1973)}]{nei:Genetics:1973}
\bibinfo{author}{\bibfnamefont{M.}~\bibnamefont{Nei}} \bibnamefont{and}
  \bibinfo{author}{\bibfnamefont{A.~A.~K.} \bibnamefont{Roychoudhury}},
  \bibinfo{journal}{Genetics} \textbf{\bibinfo{volume}{74}},
  \bibinfo{pages}{371} (\bibinfo{year}{1973}).

\bibitem[{\citenamefont{Slatkin}(1981)}]{slatkin:Evolution:1981}
\bibinfo{author}{\bibfnamefont{M.}~\bibnamefont{Slatkin}},
  \bibinfo{journal}{Evolution} \textbf{\bibinfo{volume}{35}},
  \bibinfo{pages}{477} (\bibinfo{year}{1981}).

\bibitem[{\citenamefont{Taylor et~al.}(2006)\citenamefont{Taylor, Iwasa, and
  Nowak}}]{taylor:JTB:2006}
\bibinfo{author}{\bibfnamefont{C.}~\bibnamefont{Taylor}},
  \bibinfo{author}{\bibfnamefont{Y.}~\bibnamefont{Iwasa}}, \bibnamefont{and}
  \bibinfo{author}{\bibfnamefont{M.~A.} \bibnamefont{Nowak}},
  \bibinfo{journal}{Journal of Theoretical Biology}
  \textbf{\bibinfo{volume}{243}}, \bibinfo{pages}{245} (\bibinfo{year}{2006}).

\bibitem[{\citenamefont{Altrock and Traulsen}(2009)}]{altrock:NJP:2009}
\bibinfo{author}{\bibfnamefont{P.~M.} \bibnamefont{Altrock}} \bibnamefont{and}
  \bibinfo{author}{\bibfnamefont{A.}~\bibnamefont{Traulsen}},
  \bibinfo{journal}{New Journal of Physics} \textbf{\bibinfo{volume}{11}},
  \bibinfo{pages}{013012} (\bibinfo{year}{2009}).

\bibitem[{\citenamefont{Hindersin and Traulsen}(2015)}]{hindersin:PlosCB:2015}
\bibinfo{author}{\bibfnamefont{L.}~\bibnamefont{Hindersin}} \bibnamefont{and}
  \bibinfo{author}{\bibfnamefont{A.}~\bibnamefont{Traulsen}},
  \bibinfo{journal}{PLoS Computational Biology} \textbf{\bibinfo{volume}{11}},
  \bibinfo{pages}{e1004437} (\bibinfo{year}{2015}).

\bibitem[{\citenamefont{Frean et~al.}(2013)\citenamefont{Frean, Rainey, and
  Traulsen}}]{frean:PRSB:2013}
\bibinfo{author}{\bibfnamefont{M.}~\bibnamefont{Frean}},
  \bibinfo{author}{\bibfnamefont{P.}~\bibnamefont{Rainey}}, \bibnamefont{and}
  \bibinfo{author}{\bibfnamefont{A.}~\bibnamefont{Traulsen}},
  \bibinfo{journal}{Proceedings of the Royal Society B}
  \textbf{\bibinfo{volume}{280}}, \bibinfo{pages}{20130211}
  (\bibinfo{year}{2013}).

\bibitem[{\citenamefont{Ohtsuki et~al.}(2006)\citenamefont{Ohtsuki, Hauert,
  Lieberman, and Nowak}}]{ohtsuki:Nature:2006}
\bibinfo{author}{\bibfnamefont{H.}~\bibnamefont{Ohtsuki}},
  \bibinfo{author}{\bibfnamefont{C.}~\bibnamefont{Hauert}},
  \bibinfo{author}{\bibfnamefont{E.}~\bibnamefont{Lieberman}},
  \bibnamefont{and} \bibinfo{author}{\bibfnamefont{M.~A.} \bibnamefont{Nowak}},
  \bibinfo{journal}{Nature} \textbf{\bibinfo{volume}{441}},
  \bibinfo{pages}{502} (\bibinfo{year}{2006}).

\bibitem[{\citenamefont{Ohtsuki and
  Nowak}(2006{\natexlab{b}})}]{ohtsuki:PRSB:2006}
\bibinfo{author}{\bibfnamefont{H.}~\bibnamefont{Ohtsuki}} \bibnamefont{and}
  \bibinfo{author}{\bibfnamefont{M.~A.} \bibnamefont{Nowak}},
  \bibinfo{journal}{Proceedings of the Royal Society B}
  \textbf{\bibinfo{volume}{273}}, \bibinfo{pages}{2249}
  (\bibinfo{year}{2006}{\natexlab{b}}).

\bibitem[{\citenamefont{Karlin and Taylor}(1975)}]{karlin:book:1975}
\bibinfo{author}{\bibfnamefont{S.}~\bibnamefont{Karlin}} \bibnamefont{and}
  \bibinfo{author}{\bibfnamefont{H.~M.~A.} \bibnamefont{Taylor}},
  \emph{\bibinfo{title}{A First Course in Stochastic Processes}}
  (\bibinfo{publisher}{Academic}, \bibinfo{address}{London},
  \bibinfo{year}{1975}), \bibinfo{edition}{2nd} ed.

\bibitem[{\citenamefont{Altrock et~al.}(2012)\citenamefont{Altrock, Traulsen,
  and Galla}}]{altrock:JTB:2012}
\bibinfo{author}{\bibfnamefont{P.~M.} \bibnamefont{Altrock}},
  \bibinfo{author}{\bibfnamefont{A.}~\bibnamefont{Traulsen}}, \bibnamefont{and}
  \bibinfo{author}{\bibfnamefont{T.}~\bibnamefont{Galla}},
  \bibinfo{journal}{Journal of Theoretical Biology}
  \textbf{\bibinfo{volume}{311}}, \bibinfo{pages}{94} (\bibinfo{year}{2012}).

\bibitem[{\citenamefont{Ewens}(2004)}]{ewens:book:2004}
\bibinfo{author}{\bibfnamefont{W.~J.} \bibnamefont{Ewens}},
  \emph{\bibinfo{title}{Mathematical Population Genetics. I. Theoretical
  Introduction}} (\bibinfo{publisher}{Springer}, \bibinfo{address}{New York},
  \bibinfo{year}{2004}).

\bibitem[{\citenamefont{Nowak and Sigmund}(2004)}]{nowak:Science:2004}
\bibinfo{author}{\bibfnamefont{M.~A.} \bibnamefont{Nowak}} \bibnamefont{and}
  \bibinfo{author}{\bibfnamefont{K.}~\bibnamefont{Sigmund}},
  \bibinfo{journal}{Science} \textbf{\bibinfo{volume}{303}},
  \bibinfo{pages}{793} (\bibinfo{year}{2004}).

\bibitem[{\citenamefont{Wu et~al.}(2010)\citenamefont{Wu, Altrock, Wang, and
  Traulsen}}]{wu:PRE:2010}
\bibinfo{author}{\bibfnamefont{B.}~\bibnamefont{Wu}},
  \bibinfo{author}{\bibfnamefont{P.~M.} \bibnamefont{Altrock}},
  \bibinfo{author}{\bibfnamefont{L.}~\bibnamefont{Wang}}, \bibnamefont{and}
  \bibinfo{author}{\bibfnamefont{A.}~\bibnamefont{Traulsen}},
  \bibinfo{journal}{Physical Review E} \textbf{\bibinfo{volume}{82}},
  \bibinfo{pages}{046106} (\bibinfo{year}{2010}).

\bibitem[{\citenamefont{Wu et~al.}(2013)\citenamefont{Wu, Garc{\'\i}a, Hauert,
  and Traulsen}}]{wu:PlosCB:2013}
\bibinfo{author}{\bibfnamefont{B.}~\bibnamefont{Wu}},
  \bibinfo{author}{\bibfnamefont{J.}~\bibnamefont{Garc{\'\i}a}},
  \bibinfo{author}{\bibfnamefont{C.}~\bibnamefont{Hauert}}, \bibnamefont{and}
  \bibinfo{author}{\bibfnamefont{A.}~\bibnamefont{Traulsen}},
  \bibinfo{journal}{PLoS Computational Biology} \textbf{\bibinfo{volume}{9}},
  \bibinfo{pages}{e1003381} (\bibinfo{year}{2013}).

\bibitem[{\citenamefont{Wu et~al.}(2015)\citenamefont{Wu, Bauer, Galla, and
  Traulsen}}]{wu:NJP:2015}
\bibinfo{author}{\bibfnamefont{B.}~\bibnamefont{Wu}},
  \bibinfo{author}{\bibfnamefont{B.}~\bibnamefont{Bauer}},
  \bibinfo{author}{\bibfnamefont{T.}~\bibnamefont{Galla}}, \bibnamefont{and}
  \bibinfo{author}{\bibfnamefont{A.}~\bibnamefont{Traulsen}},
  \bibinfo{journal}{New Journal of Physics} \textbf{\bibinfo{volume}{17}},
  \bibinfo{pages}{023043} (\bibinfo{year}{2015}).

\bibitem[{\citenamefont{Traulsen et~al.}(2008)\citenamefont{Traulsen, Shoresh,
  and Nowak}}]{traulsen:bmb:2008}
\bibinfo{author}{\bibfnamefont{A.}~\bibnamefont{Traulsen}},
  \bibinfo{author}{\bibfnamefont{N.}~\bibnamefont{Shoresh}}, \bibnamefont{and}
  \bibinfo{author}{\bibfnamefont{M.~A.} \bibnamefont{Nowak}},
  \bibinfo{journal}{Bulletin of Mathematical Biology}
  \textbf{\bibinfo{volume}{70}}, \bibinfo{pages}{1410} (\bibinfo{year}{2008}).

\bibitem[{\citenamefont{Antal et~al.}(2006)\citenamefont{Antal, Redner, and
  Sood}}]{antal:PRL:2006}
\bibinfo{author}{\bibfnamefont{T.}~\bibnamefont{Antal}},
  \bibinfo{author}{\bibfnamefont{S.}~\bibnamefont{Redner}}, \bibnamefont{and}
  \bibinfo{author}{\bibfnamefont{V.}~\bibnamefont{Sood}},
  \bibinfo{journal}{Physical Review Letters} \textbf{\bibinfo{volume}{96}},
  \bibinfo{pages}{188104} (\bibinfo{year}{2006}).

\bibitem[{\citenamefont{Szab{\'o} and F{\'a}th}(2007)}]{szabo:PR:2007}
\bibinfo{author}{\bibfnamefont{G.}~\bibnamefont{Szab{\'o}}} \bibnamefont{and}
  \bibinfo{author}{\bibfnamefont{G.}~\bibnamefont{F{\'a}th}},
  \bibinfo{journal}{Physics Reports} \textbf{\bibinfo{volume}{446}},
  \bibinfo{pages}{97} (\bibinfo{year}{2007}).

\bibitem[{\citenamefont{Zukewich et~al.}(2013)\citenamefont{Zukewich, Kurella,
  Doebeli, and Hauert}}]{zukewich:PlosOne:2013}
\bibinfo{author}{\bibfnamefont{J.}~\bibnamefont{Zukewich}},
  \bibinfo{author}{\bibfnamefont{V.}~\bibnamefont{Kurella}},
  \bibinfo{author}{\bibfnamefont{M.}~\bibnamefont{Doebeli}}, \bibnamefont{and}
  \bibinfo{author}{\bibfnamefont{C.}~\bibnamefont{Hauert}},
  \bibinfo{journal}{PLoS One} \textbf{\bibinfo{volume}{8}},
  \bibinfo{pages}{e54639} (\bibinfo{year}{2013}).

\bibitem[{\citenamefont{Graham et~al.}(1994)\citenamefont{Graham, Knuth, and
  Patashnik}}]{graham:book:1994}
\bibinfo{author}{\bibfnamefont{R.~L.} \bibnamefont{Graham}},
  \bibinfo{author}{\bibfnamefont{D.~E.} \bibnamefont{Knuth}}, \bibnamefont{and}
  \bibinfo{author}{\bibfnamefont{O.}~\bibnamefont{Patashnik}},
  \emph{\bibinfo{title}{Concrete Mathematics}}
  (\bibinfo{publisher}{Addison-Wesley}, \bibinfo{year}{1994}),
  \bibinfo{edition}{2nd} ed.

\bibitem[{\citenamefont{Traulsen et~al.}(2006)\citenamefont{Traulsen, Pacheco,
  and Imhof}}]{traulsen:PRE:2006c}
\bibinfo{author}{\bibfnamefont{A.}~\bibnamefont{Traulsen}},
  \bibinfo{author}{\bibfnamefont{J.~M.} \bibnamefont{Pacheco}},
  \bibnamefont{and} \bibinfo{author}{\bibfnamefont{L.~A.} \bibnamefont{Imhof}},
  \bibinfo{journal}{Physical Review E} \textbf{\bibinfo{volume}{74}},
  \bibinfo{pages}{021905} (\bibinfo{year}{2006}).

\bibitem[{\citenamefont{Cooper}(1998)}]{cooper:book:1998}
\bibinfo{author}{\bibfnamefont{R.}~\bibnamefont{Cooper}},
  \emph{\bibinfo{title}{Coordination Games}} (\bibinfo{publisher}{Cambridge
  University Press, Cambridge, UK}, \bibinfo{year}{1998}).

\bibitem[{\citenamefont{Milinski et~al.}(2008)\citenamefont{Milinski,
  Sommerfeld, Krambeck, Reed, and Marotzke}}]{milinski:PNAS:2008}
\bibinfo{author}{\bibfnamefont{M.}~\bibnamefont{Milinski}},
  \bibinfo{author}{\bibfnamefont{R.~D.} \bibnamefont{Sommerfeld}},
  \bibinfo{author}{\bibfnamefont{H.-J.} \bibnamefont{Krambeck}},
  \bibinfo{author}{\bibfnamefont{F.~A.} \bibnamefont{Reed}}, \bibnamefont{and}
  \bibinfo{author}{\bibfnamefont{J.}~\bibnamefont{Marotzke}},
  \bibinfo{journal}{Proceedings of the National Academy of Sciences USA}
  \textbf{\bibinfo{volume}{105}}, \bibinfo{pages}{2291} (\bibinfo{year}{2008}).

\bibitem[{\citenamefont{Hilbe et~al.}(2013)\citenamefont{Hilbe, Abou~Chakra,
  Altrock, and Traulsen}}]{hilbe:PlosOne:2013a}
\bibinfo{author}{\bibfnamefont{C.}~\bibnamefont{Hilbe}},
  \bibinfo{author}{\bibfnamefont{M.}~\bibnamefont{Abou~Chakra}},
  \bibinfo{author}{\bibfnamefont{P.~M.} \bibnamefont{Altrock}},
  \bibnamefont{and} \bibinfo{author}{\bibfnamefont{A.}~\bibnamefont{Traulsen}},
  \bibinfo{journal}{PLoS One} \textbf{\bibinfo{volume}{6}},
  \bibinfo{pages}{e66490} (\bibinfo{year}{2013}).

\bibitem[{\citenamefont{Altrock et~al.}(2011)\citenamefont{Altrock, Traulsen,
  and Reed}}]{altrock:PLoSCB:2011}
\bibinfo{author}{\bibfnamefont{P.~M.} \bibnamefont{Altrock}},
  \bibinfo{author}{\bibfnamefont{A.}~\bibnamefont{Traulsen}}, \bibnamefont{and}
  \bibinfo{author}{\bibfnamefont{F.~A.} \bibnamefont{Reed}},
  \bibinfo{journal}{PLoS Computational Biology} \textbf{\bibinfo{volume}{7}},
  \bibinfo{pages}{e1002260} (\bibinfo{year}{2011}).

\bibitem[{\citenamefont{MacLean et~al.}(2010)\citenamefont{MacLean,
  Fuentes-Hernandez, Greig, Hurst, and Gudelj}}]{maclean:PloSB:2010}
\bibinfo{author}{\bibfnamefont{G.}~\bibnamefont{MacLean}},
  \bibinfo{author}{\bibfnamefont{A.}~\bibnamefont{Fuentes-Hernandez}},
  \bibinfo{author}{\bibfnamefont{D.}~\bibnamefont{Greig}},
  \bibinfo{author}{\bibfnamefont{L.~D.} \bibnamefont{Hurst}}, \bibnamefont{and}
  \bibinfo{author}{\bibfnamefont{I.}~\bibnamefont{Gudelj}},
  \bibinfo{journal}{PLoS Biology} \textbf{\bibinfo{volume}{8}},
  \bibinfo{pages}{e1000486} (\bibinfo{year}{2010}).

\bibitem[{\citenamefont{Doebeli and Hauert}(2005)}]{doebeli:EL:2005}
\bibinfo{author}{\bibfnamefont{M.}~\bibnamefont{Doebeli}} \bibnamefont{and}
  \bibinfo{author}{\bibfnamefont{C.}~\bibnamefont{Hauert}},
  \bibinfo{journal}{Ecology Letters} \textbf{\bibinfo{volume}{8}},
  \bibinfo{pages}{748} (\bibinfo{year}{2005}).

\bibitem[{\citenamefont{Antal and Scheuring}(2006)}]{antal:BMB:2006}
\bibinfo{author}{\bibfnamefont{T.}~\bibnamefont{Antal}} \bibnamefont{and}
  \bibinfo{author}{\bibfnamefont{I.}~\bibnamefont{Scheuring}},
  \bibinfo{journal}{Bulletin of Mathematical Biology}
  \textbf{\bibinfo{volume}{68}}, \bibinfo{pages}{1923} (\bibinfo{year}{2006}).

\bibitem[{\citenamefont{Maciejewski et~al.}(2014)\citenamefont{Maciejewski, Fu,
  and Hauert}}]{maciejewski:PLoSCB:2014}
\bibinfo{author}{\bibfnamefont{W.}~\bibnamefont{Maciejewski}},
  \bibinfo{author}{\bibfnamefont{F.}~\bibnamefont{Fu}}, \bibnamefont{and}
  \bibinfo{author}{\bibfnamefont{C.}~\bibnamefont{Hauert}},
  \bibinfo{journal}{PLoS Computational Biology} \textbf{\bibinfo{volume}{10}},
  \bibinfo{pages}{e1003567} (\bibinfo{year}{2014}).

\bibitem[{\citenamefont{Goel and Richter-Dyn}(1974)}]{goel:book:1974}
\bibinfo{author}{\bibfnamefont{N.}~\bibnamefont{Goel}} \bibnamefont{and}
  \bibinfo{author}{\bibfnamefont{N.}~\bibnamefont{Richter-Dyn}},
  \emph{\bibinfo{title}{Stochastic Models in Biology}}
  (\bibinfo{publisher}{Academic Press, New York}, \bibinfo{year}{1974}).

\bibitem[{\citenamefont{Askari and Samani}(2015)}]{askari:PRE:2015}
\bibinfo{author}{\bibfnamefont{M.}~\bibnamefont{Askari}} \bibnamefont{and}
  \bibinfo{author}{\bibfnamefont{K.~A.} \bibnamefont{Samani}},
  \bibinfo{journal}{Physical Review E} \textbf{\bibinfo{volume}{92}},
  \bibinfo{pages}{042707} (\bibinfo{year}{2015}).

\end{thebibliography}
\end{document}